\PassOptionsToPackage{unicode}{hyperref}

\documentclass[floatfix,aps,prd,amsmath,nofootinbib,onecolumn,11pt]{revtex4-1}
\def \be {\begin{equation}} 
\def \ee {\end{equation}} 
\def \bea {\begin{eqnarray}} 
\def \eea {\end{eqnarray}}

\usepackage{graphicx}
\usepackage{dcolumn}
\usepackage{bm}
\usepackage{epsfig} 
\usepackage{amsfonts}
\usepackage{amsmath}
\usepackage{amssymb}
\usepackage[table,dvipsnames]{xcolor}
\usepackage{multirow}
\usepackage{orcidlink}
\usepackage{hyperref}

\hypersetup{
    colorlinks=true,
    citecolor=blue,
    linkcolor=blue,
    urlcolor=blue
}

\begin{document}

\title{The Impact of Inclination, Redshift, and Mass Ratio Selection on Cosmological Constraints from Bright Sirens Beyond Conventional Fisher Approximations}

\author{Gustavo B. F. Lima}
\email{gustavolima@on.br}
\affiliation{Observat\'orio Nacional, 20921-400, Rio de Janeiro - RJ, Brazil}

\author{Rafael F. Aranha\orcidlink{0000-0002-9832-2379}}
\email{rafael.aranha@uerj.br} 
\affiliation{Department of Theoretical Physics, Rio de Janeiro State University\\524 S\~{a}o Francisco Xavier Street, Rio de Janeiro, RJ, Brazil 20550-900}

\author{Rodrigo S. Gonçalves}
\email{rsg\underline{ }goncalves@ufrrj.br}
\affiliation{Departamento de Física, Universidade Federal Rural do Rio de Janeiro, 23897-000, Seropédica - RJ, Brazil}
\affiliation{Observat\'orio Nacional, 20921-400, Rio de Janeiro - RJ, Brazil}

\date{\today}

\begin{abstract}
Future generations of gravitational-wave observatories are projected to detect up to thousands bright siren events within a decade of operation. However, the precision of these cosmological forecasts depends heavily on the astrophysical assumptions embedded within the simulation pipelines. In this work, we systematically investigate the impact of variations in the binary mass ratio ($q$), the maximum observable inclination angle ($\iota_{\max}$), and the maximum redshift ($z_{\max}$) on the constraints of the Hubble constant ($H_0$) and the matter density parameter ($\Omega_m$). We perform our analyses based on the GWDALI framework, the GW Doublet Approximation, and the phenomenological waveform approximant IMRPhenomD. By adopting sets of non-spinning binary neutron star populations we explore the impact of astrophysical parameters on the constraints of cosmological parameters. Our results reveal a clear dichotomy in the observational requirements: while the statistical uncertainty on $H_0$ falls below the falls below approximately $1.3\%$ level for $N \ge 400$ events and drops under $0.9\%$ once $N \ge 800$, a deep-redshift catalog ($z_{\max} \ge 1.0$) is strictly required to effectively break parameter degeneracies and constrain the global geometry through $\Omega_m$. Furthermore, expanding the maximum inclination cutoff from the conservative $20^{\circ}$ up to $40^{\circ}$ substantially mitigates the correlation between distance and inclination.
\end{abstract}

\maketitle

\section{Introduction}
\label{sec:intro}

The first direct detection of a gravitational wave signal occurred in 2015, achieved by the LIGO collaboration \cite{Abbott:2016blz}. Since then, standard siren cosmology has become a well-established observational tool, with approximately 250 candidate gravitational-wave signals reported in real time during the fourth observing run (O4) of the LIGO/Virgo/KAGRA collaboration \cite{LVK_O4_250}. The majority of these events are classified as dark sirens, for which no electromagnetic counterpart has been identified \cite{Abbott:2021djp}, \cite{Abbott:2019yzh}. Considerable effort has been dedicated to extracting cosmological information from such events, despite the absence of precise sky localization and redshift measurements \cite{Schutz1986_HubbleGW, Muttoni_2023_dark_sirens, Alfradique_2023_dark_siren_measurement}. Some approaches employ statistical cross-correlation with galaxy catalogs, while others rely on prior knowledge of the source mass distribution to constrain cosmological parameters \cite{Ezquiaga_Holz_2022_spectral_sirens,Farah_2024_spectral_sirens}.

Among the gravitational wave sources the so-called bright sirens are events with an observed electromagnetic counterpart, and these events are considerably more valuable for cosmological analyses, since the counterpart provides the redshift necessary to break the degeneracy between luminosity distance and redshift \cite{Souza_Sturani_2021_ET_cosmography} and \cite{Zhang2019_ET_standard_sirens}. So far, only one event of this type has been observed, GW170817 \cite{Abbott2017_GW170817}, which highlights the challenges associated with detecting and identifying multimessenger events.

Even being a single event it has been proven to be sufficient to estimate the Hubble constant, demonstrating the potential of these observables \cite{Abbott2017_H0_GW170817}. The next generation of interferometers is expected to increase the number of such detections substantially. Once the Einstein Telescope and Cosmic Explorer are operational, projections indicate up to 1000 bright siren detections within a decade of operation. Recent studies suggest that the resulting constraints could make bright sirens competitive with standard candles and even the cosmic microwave background (CMB) as cosmological probes \cite{Souza_Sturani_2021_ET_cosmography} and \cite{Zhang2019_ET_standard_sirens}.

Standard sirens, however, do not have a simple mathematical description. Accurate modeling of the gravitational wave signal requires a large number of astrophysical parameters, and numerical relativity simulations are often necessary, involving high computational cost \cite{Pretorius2005}. To address this challenge, several phenomenological waveform models, such as IMRPhenomD (abbreviated as PhenomD) \cite{Husa2016_IMRPhenomI} and \cite{Khan2016_IMRPhenomD}, have been developed. The present work focuses on understanding how astrophysical parameters influence cosmological inference, particularly in the estimation of the Hubble constant and matter density. Special attention is given to the mass ratio parameter, which remains one of the most challenging quantities to model in numerical relativity.

Another aspect investigated in this study concerns the maximum inclination angle and redshift at which electromagnetic counterparts can still be detected. Recent observations and theoretical studies indicate that short gamma-ray bursts (sGRBs) associated with binary neutron star (BNS) mergers can remain detectable for inclination angles larger than the traditionally assumed limit of $20^\circ$. In particular, structured jet models can produce significant off-axis emission, extending the possible detection range to inclination angles of up to approximately $40^\circ$ \cite{Saleem2019_inclination_distribution,Kathirgamaraju_2017,Ghirlanda_2019,Mooley_2018}. The present analysis also explores how different assumptions regarding the maximum observable redshift affect the estimation of cosmological parameters, given the corresponding change in the number of detectable bright sirens.

For the simulations and analyses presented in this work, the GWDALI framework is employed \cite{de_Souza_2023}. It is designed to explore gravitational-wave populations and their cosmological implications through a Bayesian approach. Unlike tools such as GWFISH \cite{Dupletsa_2023}, which rely primarily on the Fisher-matrix formalism, GWDALI allows for a more flexible treatment of parameter estimation. The Fisher approach can become unreliable when the likelihood deviates from Gaussianity or when strong parameter degeneracies are present, which may lead to instabilities in the inversion of the Fisher matrix and consequently to inaccurate uncertainty estimates. By incorporating more realistic observational uncertainties and population models, GWDALI provides a more robust framework to evaluate how astrophysical parameters, such as the mass ratio and inclination, propagate into cosmological constraints.





\section{Phenomenological Waveform Model for Non-Precessing Black Hole Binaries}
\label{sec:phenommodel}

\subsection{Overview: Numerical, Post-Newtonian, and Hybrid Waveforms}

The gravitational waveform emitted by a coalescing binary black hole system encodes detailed information about the masses, spins, and orbital dynamics of the system. In the frequency ($f$) domain, the complex strain amplitude for a given spherical harmonic $(1=2,|m|=2)$-mode can be written as

\begin{equation}
\tilde{h}(f,\Xi) = A(f,\Xi) e^{i\Phi(f,\Xi)} \, ,
\label{eq:strain}
\end{equation}
where $A(f,\Xi)$ and $\Phi(f,\Xi)$ are respectively the amplitude and the phase of the wave and $\Xi \in (M,\eta,\chi_1 , \chi_2)$ are the parameters of the binary consisted of masses $m_1$ and $m_2$: the total mass $M=m_1 + m_2$, the symmetric mass ratio $\eta=m_1 m_2/M^2$ and black holes spins $\chi_1$ and $\chi_2$. The amplitude and phase fully characterize the waveform as they contain complementary information: the phase governs the timing of the inspiral and is crucial for matched-filtering detection, while the amplitude determines the signal strength and also influences parameter estimation.

In general, three complementary approaches are used to model these waveforms across the full frequency band relevant to ground-based detectors, as follows.

\subsubsection{Post-Newtonian Waveforms}

The post-Newtonian (PN) formalism provides an analytical description of the early inspiral under the assumption of weak fields and slow velocities, expanding all physical quantities in powers of $v/c$. For non-precessing systems with spins aligned to the orbital angular momentum, the TaylorF2 approximant \cite{taylorTF2-1,taylorTF2-2,taylorTF2-3,taylorTF2-4} gives the frequency-domain phase and amplitude in a closed form. The phase is derived using the stationary phase approximation, including spin-orbit and spin-spin contributions,

\begin{equation}
\Phi_{\mathrm{TF2}}(Mf) = 2\pi f t_0 - \phi_0 - \frac{\pi}{4} + \frac{3}{128\eta} (\pi Mf)^{-5/3} \sum_{k=0}^{7} \varphi_k (\Xi) (\pi Mf)^{k/3} \, ,
\label{eq:pn_phase}
\end{equation}
where $t_0$ and $\phi_0$ are integration constants from the PN procedure. The coefficients $\varphi_k(\Xi)$ (which incorporate spin effects) are given in the appendix B of \cite{Khan2016_IMRPhenomD}.

The PN amplitude can be written as

\begin{equation}
A_{\mathrm{PN}}(f) = \sqrt{\frac{2\eta}{3\pi^{1/3}}}f^{-7/6} \left(\sum_{k=0}^{6} \mathcal{A}_k(f,\chi_1 , \chi_2) (Mf)^{k/2}\right) ,
\label{eq:pn_amplitude}
\end{equation}
where the expansion coefficients $\mathcal{A}_k(f,\chi_1 , \chi_2)$ are also obtained in the appendix B of \cite{Khan:2015jqa}. While PN waveforms are accurate at low frequencies, they become unreliable as the binary approaches merger, where the expansion parameter $v/c$ ceases to be small. Then, a full general relativistic numerical scheme must play a leading role.

\subsubsection{Numerical Relativity Waveforms}
\label{Sec:NumRelWave}

Numerical relativity (NR) solves the full Einstein field equations without approximation, providing waveforms that are exact within numerical error. For binary black holes, NR simulations cover the late inspiral, merger, and ringdown phases. The waveforms are typically extracted from the Weyl scalar $\Psi_4$, which is related to the strain by time second derivatives of the gravitational waves polarizations as

\begin{equation}
\Psi_4 = \frac{d^2}{dt^2} (h_+ - i h_\times)=\sum_{m=-2,2}{\psi}_{2m}(t,\Xi) ^{-2}Y_{2m}(\theta,\phi),
\label{eq:psi4}
\end{equation}

\noindent where $^{-2}Y_{2m}(\theta,\phi)$ are the $l=2, |m|=2$ spin weighted spherical harmonics, ${\psi}_{2m}(t,\Xi)$ are the expansion coefficients of the wave signal and $\psi_{2-2}(f)=\psi^{*}_{22}(-f)$. In the frequency domain, the strain coefficient $\tilde{h}_{22}(f,\Xi)$ provides the amplitude and the phase of the numerical relativity gravitational wave, 

\begin{eqnarray}
\tilde{h}_{\mathrm{NR}}(f,\Xi) \equiv \tilde{h}_{22}(f,\Xi) = -\frac{\tilde{\psi}_{22}(f,\Xi)}{4\pi^2 f^2} = A_{\mathrm{NR}}(f,\Xi) e^{i\Phi_{\mathrm{NR}}(f,\Xi)}.
\label{eq:nr_strain}
\end{eqnarray}

NR waveforms are accurate at high frequencies but are computationally expensive to produce, limiting the number of orbits that can be simulated. The NR waveforms from the SXS collaboration and new BAM simulations \cite{sxs,bam1,bam2} cover mass ratios up to $\tilde{q} = {m_2}/{m_1} = 18$ and spins up to $|\chi_i| \leq 0.98$ for equal-mass binaries. 
The simulations consider two sources of error: errors due to i) finite numerical resolution and ii) finite waveform extraction radius. In all cases the overall mismatch error from these sources are less than $0.5\%$.

\subsubsection{Hybrid Waveforms}

To obtain a complete waveform spanning from low to high frequencies, PN and NR waveforms are combined through a matching procedure. In such hybrid construction, the inspiral is described by the uncalibrated SEOBv2 (Spinning Effective One-Body, version 2) model that is the effective-one-body approximant without NR-fitted corrections which has been shown to be more consistent with NR data than standard PN approximants \cite{taracchini}.

The hybrid waveform is constructed as

\begin{equation}
\tilde{h}_{\mathrm{hybrid}}(f) = \begin{cases}
\tilde{h}_{\mathrm{SEOBv2}}(f), & f \leq f_{\mathrm{match}},\\
\tilde{h}_{\mathrm{NR}}(f), & f > f_{\mathrm{match}} \, ,
\end{cases}
\vspace{0,5cm}
\label{eq:hybrid}
\end{equation}
with a smooth transition in the matching window. The construction requires careful alignment of the phase and amplitude between the two descriptions, accounting for systematic differences in parameter definitions and gauge choices. The resulting hybrid waveforms serve as benchmark models against which phenomenological approximations can be calibrated. See Refs. \cite{hyb1,hyb2,hyb3,hyb4} for summaries of different methods for hybrid waveform modeling in the literature.

\subsection{The Phenomenological Strategy}

The phenomenological modeling approach aims to replace the expensive and sparse set of hybrid waveforms with a simple analytical formula that can be evaluated rapidly for any parameter combination. Throughout the text, we employ the phenomenological methodology associated with the PhenomD waveform catalog. It is also common in the literature to refer to the catalog as IMRPhenomD, in order to explicitly indicate that the waveforms encompass the inspiral, merger, and ringdown phases of binary black hole coalescence. The key observation is that the waveform is fully characterized by its amplitude and phase. Moreover, these two quantities exhibit distinct physical features which can be modeled separately:
\begin{itemize}
    \item[i)] \textit{Phase:} Dominated by the orbital dynamics, the phase accumulates over many cycles and is highly sensitive to the system's parameters. It determines the signal's timing and is the primary quantity for matched-filtering detection. Physically, the phase is directly related to the orbital motion through $\Phi(t) \approx 2\int \omega(t) dt$, so modeling the phase is equivalent to modeling the binary dynamics.
    
    \item[ii)] \textit{Amplitude:} Governed by the energy flux and the system luminosity, the amplitude evolves more smoothly and influences the overall signal-to-noise ratio and parameter estimation biases. In the inspiral, it follows a well-known power law $A(f) \propto f^{-7/6}$, modulated by higher-order corrections.
\end{itemize}
\indent \par By modeling $A(f)$ and $\Phi(f)$ independently, we can leverage known analytical results from PN/EOB theory for the inspiral and from black hole perturbation theory for the ringdown, while using phenomenological functions to bridge the merger region where no complete analytical description exists. This separation (of amplitude and phase) also allows for independent control of the smoothness and accuracy of each component.

\subsection{Construction of Hybrid Waveforms for Aligned-Spin Systems}

The hybrid waveforms used in this work are constructed following the procedure of \cite{Khan:2015jqa}. The inspiral component uses the uncalibrated SEOBv2 model, which has been shown to agree better with NR data than standard PN approximants, especially for high mass ratios and spins. The NR waveforms are drawn from multiple independent codes (BAM and SpEC) covering mass ratios $m_2/m_1 \in [1, 18]$ and aligned spins $|\chi_i| \leq 0.98$ for equal mass and $|\chi_i| \leq 0.85$ for unequal masses.

A crucial simplification for non-precessing systems is that the waveform depends primarily on a single combination of the two spins, rather than on each spin individually. The effective spin parameter used in PhenomD is defined as

\begin{equation}
\chi_{\mathrm{eff}} = \frac{m_1 \chi_1 + m_2 \chi_2}{M}.
\label{eq:effective_spin}
\end{equation}

However, PN theory suggests that a better parameter is

\begin{equation}
\chi_{\mathrm{PN}} = \chi_{\mathrm{eff}} - \frac{38\eta}{113} (\chi_1 + \chi_2) \, ,
\label{eq:chi_pn}
\end{equation}
which is normalized to the range $[-1,1]$ as $\hat{\chi} = \chi_{\mathrm{PN}} / (1 - 76\eta/113)$. This parameter captures the dominant spin effect on the inspiral phase and is used throughout the phenomenological calibration.

The matching between the SEOBv2 inspiral and NR is performed at a frequency $Mf = 0.018$ \cite{Khan:2015jqa}, which is the lowest common GW frequency for which NR data is available across all configurations. For the phase, the extrinsic parameters $t_0$ and $\phi_0$ are optimized to minimize the integrated squared phase difference \cite{Santamaria:2010yb}. The hybrid phase is then defined by setting $\Phi_{\mathrm{hybrid}}(f) = \Phi_{\mathrm{SEOBv2}}(f)$ for $f \leq f_{\mathrm{match}}$ and $\Phi_{\mathrm{hybrid}}(f) = \Phi_{\mathrm{NR}}(f)$ for $f > f_{\mathrm{match}}$. For the amplitude, a similar procedure is applied.

\subsection{Phenomenological Phase Model}

The phenomenological phase model is constructed by partitioning the frequency domain into three regimes: inspiral (Region I), intermediate (Region IIa), and merger-ringdown (Region IIb). The key idea is to use known analytical results where available, and phenomenological extensions where they are not. The full IMR phase is given by

\begin{equation}
\Phi_{\mathrm{IMR}}(f) = \phi_{\mathrm{Ins}}(f)\theta_{f_1}^{-} + \theta_{f_1}^{+}\phi_{\mathrm{Int}}(f)\theta_{f_2}^{-} + \theta_{f_2}^{+}\phi_{\mathrm{MR}}(f) \, ,
\label{eq:full_phase}
\end{equation}
where the step functions are defined as

\begin{equation}
\theta(f - f_0) = \begin{cases}
-1, & f < f_0, \\
1, & f \geq f_0,
\end{cases}
\qquad
\theta_{f_0}^{\pm} = \frac{1}{2}\left[1 \pm \theta(f - f_0)\right] \, ,
\label{eq:step_functions}
\end{equation}
and the transition frequencies are $f_1 = 0.018$ and $f_2 = 0.5 f_{\mathrm{RD}}$, where $f_{\mathrm{RD}}$ is the characteristic frequency of the ringdown.

\subsubsection{Inspiral Regime (Region I)}

For frequencies below $Mf = 0.018$, the phase is modeled using the TaylorF2 approximant augmented with four additional higher-order PN terms whose coefficients are fitted to the SEOBv2+NR hybrid data. The ansatz is

\begin{equation}
\phi_{\mathrm{Ins}}(f) = \phi_{\mathrm{TF2}}(f) + \frac{1}{\eta}\left(\sigma_0 + \sigma_1 f + \frac{3}{4}\sigma_2 f^{4/3} + \frac{3}{5}\sigma_3 f^{5/3} + \frac{1}{2}\sigma_4 f^2\right) \, ,
\label{eq:inspiral_phase}
\end{equation}
where $\phi_{\mathrm{TF2}}$ is the standard TaylorF2 phase including both spin contributions given by equation (\ref{eq:pn_phase}). The additional terms capture the difference between the uncalibrated SEOBv2 model and the standard PN expansion, ensuring that the inspiral model accurately represents the hybrid data.

\subsubsection{Intermediate Regime (Region IIa)}

To bridge the gap between the end of the inspiral region ($Mf = 0.018$) and the start of the merger-ringdown model ($f = 0.5 f_{\mathrm{RD}}$), a simple ansatz for the phase derivative is used:

\begin{equation}
\eta \phi_{\mathrm{Int}}^{\prime}(f) = \beta_1 + \beta_2 f^{-1} + \beta_3 f^{-4}.
\label{eq:intermediate_phase_deriv}
\end{equation}

The behavior of the data over this frequency range is predominantly proportional to $1/f$, but at higher mass ratios and high anti-aligned spins, the additional $f^{-4}$ term is necessary for an accurate fit. The phase itself is obtained by integration:

\begin{equation}
\phi_{\mathrm{Int}}(f) = \frac{1}{\eta}\left(\beta_0 + \beta_1 f + \beta_2 \log(f) - \frac{\beta_3}{3} f^{-3}\right).
\label{eq:intermediate_phase}
\end{equation}

The constant and linear coefficients $\beta_0$ and $\beta_1$ are determined by continuity requirements with the inspiral and merger-ringdown regions. Specifically, $\beta_0$ and $\beta_1$ are fixed by matching the value and derivative of the phase at the boundaries $f = f_1$ and $f = f_2$, so they are not independent fitting parameters. Their values are derived from the inspiral and merger-ringdown models at the transition frequencies.

\subsubsection{Merger-Ringdown Regime (Region IIb)}

For frequencies above $f = 0.5 f_{\mathrm{RD}}$, the phase derivative is modeled with an ansatz that captures the characteristic dip (a localized drop or reduction in phase derivative) at the ringdown frequency. The derivative of the phase is given by

\begin{equation}
\eta \phi_{\mathrm{MR}}^{\prime}(f) = \alpha_1 + \alpha_2 f^{-2} + \alpha_3 f^{-1/4} + \frac{\alpha_4 f_{\mathrm{damp}}}{f_{\mathrm{damp}}^2 + (f - \alpha_5 f_{\mathrm{RD}})^2} \, ,
\label{eq:mr_phase_deriv}
\end{equation}
where $f_{\mathrm{damp}}$ is the damping frequency, associated with the imaginary part of the quasi-normal modes (QNMs) of the ringdown. The last term (Lorenzian) in eq. (\ref{eq:mr_phase_deriv}) comes from the Fourier transform of the usual time-domain gravitational strain,

\begin{equation}
h(t)=\exp[{2\pi(if_{RD}t - f_{damp}|t|)}], ~~~~ f_{RD}, f_{damp} \in \mathbb{R}.
\label{eq:time-dom-strain}
\end{equation}

The Lorentzian term models the dip in the phase derivative at the ringdown frequency, while the power-law terms account for the overall trend. The parameters $\alpha_4$ and $\alpha_5$ control the amplitude and location of the dip.

The merger-ringdown phase is obtained by integrating Eq.~(\ref{eq:mr_phase_deriv}):

\begin{equation}
\phi_{\mathrm{MR}}(f) = \frac{1}{\eta}\left\{\alpha_0 + \alpha_1 f - \alpha_2 f^{-1} + \frac{4}{3}\alpha_3 f^{3/4} + \alpha_4 \tan^{-1}\left(\frac{f - \alpha_5 f_{\mathrm{RD}}}{f_{\mathrm{damp}}}\right)\right\}.
\label{eq:mr_phase}
\end{equation}

The constant $\alpha_0$ and linear coefficient $\alpha_1$ are fixed by continuity with the intermediate region, similar to $\beta_0$ and $\beta_1$. Therefore, $\alpha_0$ and $\alpha_1$ are not independent fitting parameters; they are derived from the requirement that the phase and its derivative are continuous at the boundary $f = f_2$. The parameters that are actually fitted to the hybrid data are $\{\alpha_2, \alpha_3, \alpha_4, \alpha_5\}$ for the phase, along with the coefficients from the inspiral and intermediate regions.

\subsection{Phenomenological Amplitude Model}

The amplitude model follows a similar three-regime structure, though with a simpler ansatz. The full IMR amplitude is given by

\begin{equation}
A_{\mathrm{IMR}}(f) = A_{\mathrm{Ins}}(f)\theta_{f_1}^{-} + \theta_{f_1}^{+}A_{\mathrm{Int}}(f)\theta_{f_2}^{-} + \theta_{f_2}^{+}A_{\mathrm{MR}}(f),
\label{eq:full_amplitude}
\end{equation}

with transition frequencies $f_1 = 0.014$ and $f_2 = f_{\mathrm{peak}}$, where $f_{\mathrm{peak}}$ is the frequency at which the amplitude peaks.

\subsubsection{Inspiral Regime}

The inspiral amplitude is based on a re-expanded PN amplitude:

\begin{equation}
A_{\mathrm{Ins}}(f) = A_0 \sum_{i=0}^{6} \mathcal{A}_i (\pi f)^{i/3} + A_0 \sum_{i=1}^{3} \rho_i f^{(6+i)/3},
\label{eq:inspiral_amp}
\end{equation}

where $A_0 = \sqrt{2\eta/3\pi^{1/3}} f^{-7/6}$ is the leading-order Newtonian amplitude. The first sum is the standard PN amplitude, and the second sum contains higher-order terms fitted to the hybrid data.

\subsubsection{Intermediate Regime}

The intermediate amplitude is modeled using a collocation method: a fourth-order polynomial is constrained by the value and derivative of the amplitude at the boundaries ($f_1 = 0.014$ and $f_3 = f_{\mathrm{peak}}$) and by the amplitude value at the midpoint $f_2 = (f_1+f_3)/2$ read from the NR data:

\begin{equation}
A_{\mathrm{Int}}(f) = A_0 \left(\delta_0 + \delta_1 f + \delta_2 f^2 + \delta_3 f^3 + \delta_4 f^4\right).
\label{eq:intermediate_amp}
\end{equation}

The five coefficients $\delta_i$ are determined by solving the system $A_{\mathrm{Int}}(f_1)=v_1$, $A_{\mathrm{Int}}(f_2)=v_2$, $A_{\mathrm{Int}}(f_3)=v_3$, $A_{\mathrm{Int}}^{\prime}(f_1)=d_1$, and $A_{\mathrm{Int}}^{\prime}(f_3)=d_3$. As with the phase coefficients, $\delta_0$ is not a free parameter; it is determined by the collocation conditions. The values $v_1$, $d_1$ come from the inspiral model, $v_3$, $d_3$ from the merger-ringdown model, and $v_2$ is read directly from the hybrid data at the midpoint frequency. Thus, the only amplitude coefficient that requires fitting across the parameter space is $v_2$ (the amplitude value at the midpoint), along with the ringdown coefficients $\gamma_1,\gamma_2,\gamma_3$ and the inspiral coefficients $\rho_i$.

\subsubsection{Merger-Ringdown Regime}

For the ringdown, the amplitude is modeled as a Lorentzian modulated by an exponential decay factor as

\begin{equation}
A_{\mathrm{MR}}(f) = A_0 \, \gamma_1 \frac{\gamma_3 f_{\mathrm{damp}}}{(f - f_{\mathrm{RD}})^2 + (\gamma_3 f_{\mathrm{damp}})^2} e^{-\gamma_2 (f - f_{\mathrm{RD}})/(\gamma_3 f_{\mathrm{damp}})},
\label{eq:mr_amp}
\end{equation}

\noindent where $\gamma_1$ determines the overall amplitude of the ringdown, $\gamma_2$ controls the decay rate, and $\gamma_3$ adjusts the width of the Lorentzian. The peak frequency is given by

\begin{equation}
f_{\mathrm{peak}} = \left| f_{\mathrm{RD}} + \frac{f_{\mathrm{damp}}\gamma_3\left(\sqrt{1 - \gamma_2^2} - 1\right)}{\gamma_2} \right|,
\label{eq:peak_frequency}
\end{equation}

\noindent and the $\gamma$ parameters belong to the following intervals: $\gamma_1 \in [0.0024,0.0169]$, $\gamma_2 \in [0.54,1.5339]$, $\gamma_3 \in [1.25,1.36]$.

\subsection{Mapping to Physical Parameters}

The phenomenological model contains 11 amplitude coefficients and 14 phase coefficients. However, four amplitude coefficients (in the intermediate region) and four phase coefficients ($\alpha_0,\alpha_1,\beta_0,\beta_1$) are constrained analytically by continuity, leaving 17 phenomenological parameters that need to be mapped onto the physical parameter space. These are parameterized by the symmetric mass ratio $\eta$ and the reduced spin parameter $\chi_{PN}$.

Each phenomenological parameter $\Lambda^i$ is expressed as a polynomial ansatz in $\eta$ and $\chi_{\mathrm{PN}}$, expanded around $\chi_{\mathrm{PN}} = 1$:

\begin{equation}
\begin{aligned}
\Lambda^i = &\ \lambda_{00}^i + \lambda_{10}^i \eta \\
&+ (\chi_{\mathrm{PN}} - 1)\left(\lambda_{01}^i + \lambda_{11}^i \eta + \lambda_{21}^i \eta^2\right) \\
&+ (\chi_{\mathrm{PN}} - 1)^2\left(\lambda_{02}^i + \lambda_{12}^i \eta + \lambda_{22}^i \eta^2\right) \\
&+ (\chi_{\mathrm{PN}} - 1)^3\left(\lambda_{03}^i + \lambda_{13}^i \eta + \lambda_{23}^i \eta^2\right).
\end{aligned}
\label{eq:mapping}
\end{equation}

The expansion is performed around $\chi_{\mathrm{PN}} = 1$ because this is where the calibration waveforms are concentrated. The coefficients $\lambda_{jk}^i$ are determined by fitting to the 19 calibration hybrid waveforms. The values of these coefficients for each phenomenological parameter are tabulated in Appendix C of \cite{Khan:2015jqa}. This mapping allows the phenomenological model to be evaluated for any combination of total mass $M$, mass ratio $q$, and spins $\chi_1,\chi_2$ within the calibrated range.

\subsection{Validation and Accuracy}

The fidelity of the phenomenological model is assessed using the overlap integral with respect to the Advanced LIGO noise power spectral density. The match between two waveforms $h_1$ and $h_2$ is defined as

\begin{equation}
\mathcal{M}(h_1,h_2) = \max_{t_0,\phi_0} \frac{(\tilde{h}_1|\tilde{h}_2)}{\sqrt{(\tilde{h}_1|\tilde{h}_1)(\tilde{h}_2|\tilde{h}_2)}},
\label{eq:match}
\end{equation}

where the inner product is given by

\begin{equation}
\langle x | y \rangle = 4\,\mathrm{Re}\int_0^\infty \frac{\tilde{x}(f) \tilde{y}^*(f)}{S_n(f)} df.
\label{eq:inner}
\end{equation}

The mismatch is defined as $1 - \mathcal{M}$. For the 19 calibration hybrids and an additional 29 verification hybrids not used in the calibration, the mismatch is typically below $1\%$ for systems with total masses up to $\sim 350 M_\odot$ when using the aLIGO design sensitivity noise curve with a lower frequency cutoff of $10$ Hz. For the early aLIGO noise curve with a $30$ Hz cutoff, all mismatches are comfortably below $1\%$. The worst mismatches occur for high mass ratios and large spins, indicating the region of parameter space that requires the most improvement in future models.

The PhenomD model has been compared against the SEOBNRv2 model, which is calibrated to NR waveforms up to mass ratio 1:8 and spins up to $0.5$. For high spins ($\hat{\chi} \gtrsim 0.75$) and unequal masses, significant disagreements are observed, suggesting that the SEOBNRv2 model may not be accurate outside its calibration region. This highlights the importance of calibrating waveform models to NR simulations over the full parameter space of interest.


\indent \par The PhenomD phenomenological waveform model provides an analytical representation of gravitational waves from non-precessing binary black holes across the entire inspiral-merger-ringdown (IMR) frequency band, with ``IMR'' referring precisely to these three stages of the binary evolution, as demonstrated in this chapter. The model is calibrated to 19 hybrid SEOBv2+NR waveforms covering mass ratios up to $\tilde{q}=18$ and spins up to $|\chi_i| \leq 0.98$, and exhibits mismatches of typically less than $1\%$ against all 48 available hybrid waveforms. By separately modeling the amplitude and phase using SEOBv2/PN theory for the inspiral, perturbation theory for the ringdown, and phenomenological ansätze for the merger, the model achieves high accuracy across the parameter space relevant to Advanced LIGO and Virgo. The separation into amplitude and phase is natural given their distinct physical origins and roles in detection and parameter estimation. This model forms the basis for the waveform templates used in the subsequent analysis.

\nocite{Santamaria:2010yb,Ajith:2009bn,Taracchini:2012ig,Buonanno:2009zt}


\section{GW simulations}
\label{sec:data_sim}

Gravitational wave (GW) simulations play a crucial role in understanding the connection between the astrophysical properties of compact binary systems and the signals observed by interferometers. To perform such simulations, three key elements must be defined: a fiducial cosmology, the waveform model, and the instrument’s physical characteristics. In this work, we focus on the simplistic version of the flat $\Lambda$CDM model as our fiducial cosmology, so the Hubble parameter can be written as
\begin{equation}
    H^2(z) = H_0^2 \left( \Omega_m (1+z)^3 + 1 - \Omega_m \right),
\end{equation}
and we adopt the cosmological parameter values provided by the Planck collaboration \cite{2020aganin}, namely $H_0 = 67.66\,\mathrm{km/s/Mpc}$ and $\Omega_m = 0.311$. The main cosmological observable used in our simulations is the luminosity distance, which, assuming a flat Universe, is expressed as 
\begin{equation}
    D_L = c \, (1+z) \int_0^z \frac{dz'}{H(z')}.
\end{equation}
With the fiducial cosmology in place, we can generate a realistic distribution of binary neutron star (BNS) events. In this work, we adopt the conservative redshift distribution used in 
\cite{Zhao_2011}, which assumes that the merger rate follows a simple 
phenomenological model. This distribution is widely used in forecasts 
for BNS and neutron star black hole (NSBH) populations in gravitational-wave cosmology studies \cite{Zhang2019_ET_standard_sirens,Cai_2017,Jin_2022}.
\begin{equation}
    \label{dist}
    P(z) \propto \frac{4 \pi d_L^2 R(z)}{H(z)(1+z)^3}, \quad 
    R(z) = \begin{cases}
    1 + 2z, & z \leq 1, \\
    \frac{3}{4}(5 - z), & 1 < z < 5, \\
    0, & z \geq 5.
    \end{cases}
\end{equation}

The PhenomD model, introduced in the previous section, describes the gravitational-wave signal as an expansion in spin-weighted spherical harmonics. The corresponding frequency-domain waveform is given by
\begin{equation}
    \label{waveform}
    \tilde{h}(f;\Xi)
    = \tilde{h}_+(f;\Xi,\iota,\phi)
    - i \tilde{h}_\times(f;\Xi,\iota,\phi)
    = \sum_{m=-2}^{2} \tilde{h}_{2m}(f;\Xi)\,{}^{-2}Y_{2m}(\iota,\phi),
\end{equation}
where the strain was introduced in section~\ref{sec:phenommodel} through equation~\eqref{eq:strain}. From this expression, the two polarization modes can be written as

\begin{align}
\tilde{h}_{+}(f;\Xi,\iota)
&=
\frac{1+\cos^{2}\iota}{2}\,
A(f;\Xi)\,
e^{-i\Phi(f;\Xi)},
\\
\tilde{h}_{\times}(f;\Xi,\iota)
&=
i\cos\iota\,
A(f;\Xi)\,
e^{-i\Phi(f;\Xi)}.
\end{align}
between the orbital angular momentum of the binary and the observer's line of sight. 

Here, $q = m_1/m_2 \leq 1$ is the mass ratio, then based on the values presented in the previous section, the calibration limit of the model corresponds to $q = 1/18 \approx 0.056$. Also, in this work, we consider non-spinning binaries, i.e., $ \chi_1 = \chi_2 = 0 $, such that the total angular momentum is dominated by the orbital angular momentum. As a result, the inclination angle $ \iota $ fully characterizes the orientation of the system with respect to the observer.

The strain measured by an interferometer depends not only on the waveform itself, but also on the geometry of the detector and the relative orientation of the source. The measurable strain $ \tilde{h}(t) $ is obtained by projecting the gravitational wave strain tensor onto the detector’s response tensor $\mathcal{P}^{ij} = \frac{1}{2} \left( e_1^i e_1^j - e_2^i e_2^j\right)$,
where $ \vec{e}_1 $ and $ \vec{e}_2 $ are unit vectors aligned with the interferometer arms. In practice, the strain can be expressed as a linear combination of the two gravitational wave polarizations modulated by the detector’s antenna pattern functions,
\begin{equation}
    h = F_+(\theta, \phi, \psi)\, \tilde{h}_+ + F_\times(\theta, \phi, \psi)\, \tilde{h}_\times,
\end{equation}
where \( (\theta, \phi) \) specify the sky location of the source, and \( \psi \) is the polarization angle.

For L-shaped interferometers, such as Cosmic Explorer, the antenna pattern functions take the form
\begin{align}
    F_+^L(\theta, \phi, \psi) &= \frac{1}{2} (1 + \cos^2\theta)\, \cos(2\phi)\, \cos(2\psi) - \cos\theta\, \sin(2\phi)\, \sin(2\psi), \\
    F_\times^L(\theta, \phi, \psi) &= \frac{1}{2} (1 + \cos^2\theta)\, \cos(2\phi)\, \sin(2\psi) + \cos\theta\, \sin(2\phi)\, \cos(2\psi).
\end{align}

For the triangular Einstein Telescope (ET) configuration, each of the three constituent interferometers has an arm opening angle of $60^\circ$. Relative to a right-angle interferometer with the same bisector orientation, this introduces a geometric factor $\sin 60^\circ = \sqrt{3}/2$ in the antenna pattern functions \cite{de_Souza_2023_luminosity_distance}:
\begin{equation}
F_+^{\triangle} = \frac{\sqrt{3}}{2} F_+^L,
\qquad
F_\times^{\triangle} = \frac{\sqrt{3}}{2} F_\times^L.
\end{equation}
Here, $F_+^{\triangle}$ and $F_\times^{\triangle}$ denote the antenna pattern functions of an individual ET interferometer. The antenna patterns are evaluated separately for the orientation of each of the three interferometers, whose contributions are then combined in the network analysis.

With all these elements in place, we can now compute the
signal-to-noise ratio (SNR) of a gravitational-wave event.
Using the inner product defined in Eq.~\eqref{eq:inner}, we obtain
\begin{equation}
\rho^2 = \langle h \mid h \rangle
= 4 \int_{f_{\mathrm{low}}}^{f_{\mathrm{upper}}}
\frac{|h(f)|^2}{S_n(f)}\,df,
\end{equation}
where $f_{\mathrm{low}}$ and $f_{\mathrm{upper}}$ are the lower
and upper integration limits, respectively. We use
$f_{\mathrm{low}}=1\,\mathrm{Hz}$ for all instruments and choose
$f_{\mathrm{upper}}$ sufficiently high to cover the frequency
range of all simulated signals.

The combined SNR from a network of interferometers is given by
\begin{equation}
\rho_{\text{net}} = \sqrt{\sum_i \rho_i^2},
\end{equation}
where $\rho_i$ denotes the SNR of the $i$-th detector. In order to retain only sufficiently significant detections, we impose a network detection threshold of $\rho_{\mathrm{net}} \geq 8$. For each detected event, the instrumental uncertainty in the luminosity distance, $\sigma_{d_L}^{\mathrm{inst}}$, is estimated using the GW Doublet Approximation implemented in GWDALI, which accounts for parameter correlations and higher-order contributions beyond the standard Fisher-matrix approximation. Once the instrumental uncertainty is computed for each event, the simulated observed luminosity distances $d_L^{\mathrm{sim}}$ are generated using truncated Monte Carlo sampling. In this approach, the fiducial distance $d_L^{\mathrm{fid}}$ is perturbed by drawing random deviations from a Gaussian distribution centered at $d_L^{\mathrm{fid}}$ with standard deviation $\sigma_{\mathrm{inst}}$, truncated to only allow positive values.

\begin{figure}[ht]
    \centering
    \includegraphics[width=0.8\linewidth]{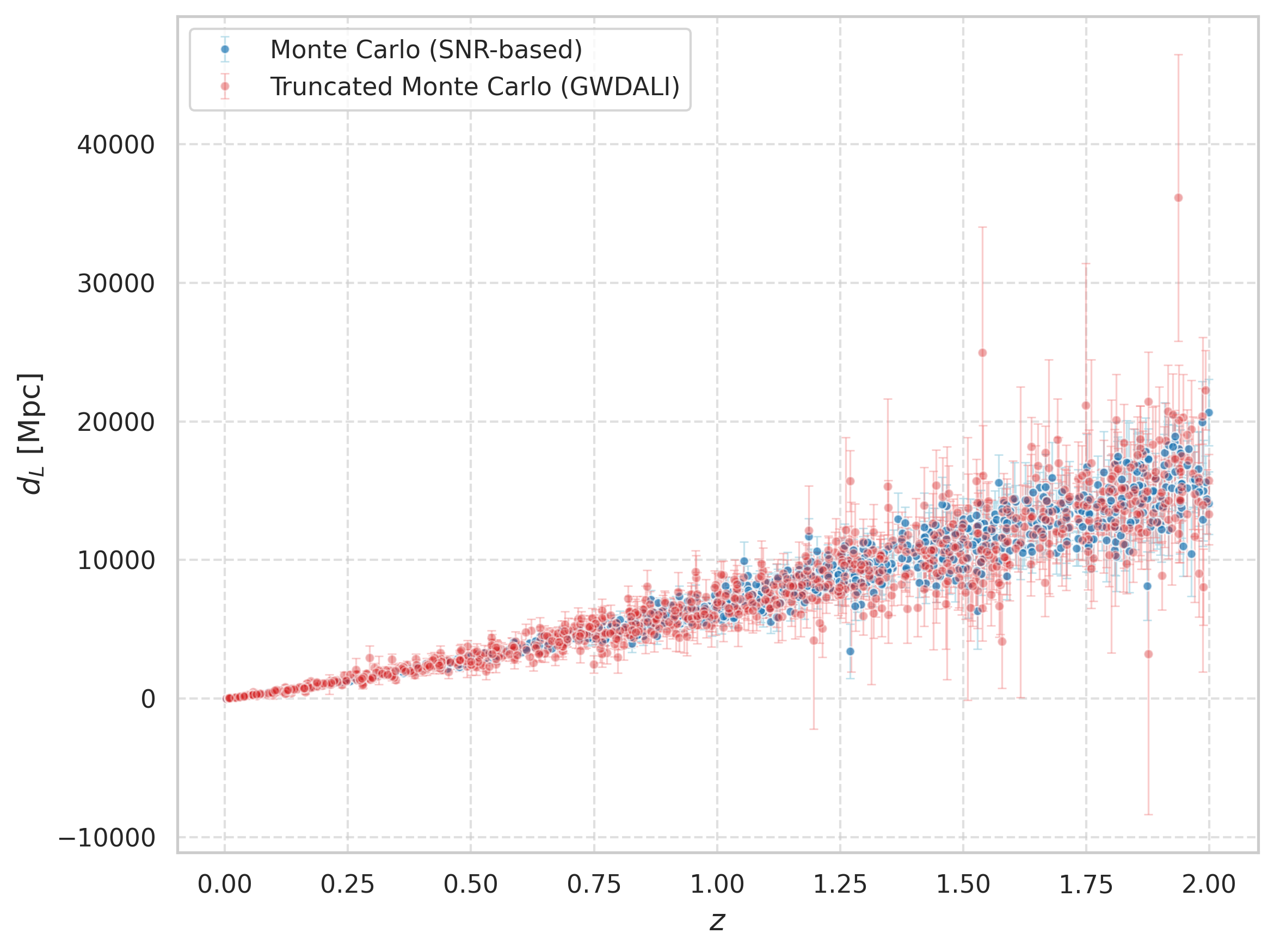}
    \caption{
Simulated luminosity distances as a function of redshift obtained using the SNR-based approximation (blue points) and the GW Doublet Approximation implemented in GWDALI (red points). The error bars represent the corresponding luminosity-distance uncertainties, while the simulated values are generated through Monte Carlo sampling around the fiducial luminosity distance.
}
    \label{fig:luminosity_distance}
\end{figure}

To ensure physically meaningful realizations, the sampling is truncated at $d_L^{\mathrm{sim}}>0$. This procedure preserves the measurement scatter predicted by GWDALI while preventing unphysical negative luminosity distances. An example of the resulting simulated catalog is shown in Fig.~\ref{fig:luminosity_distance}. For comparison, we adopt the approximation
\begin{equation}
    \sigma_{d_L} = \frac{2\,d_L}{\rho_{\mathrm{net}}},
    \label{eq:dl_error}
\end{equation}
which we refer to as the SNR-based method in Fig.~\ref{fig:luminosity_distance}. This prescription provides a simplified estimate of the instrumental distance uncertainty based directly on the network SNR. The factor of two is commonly introduced to approximately account for the degeneracy between luminosity distance and binary inclination, which is not explicitly captured when the distance uncertainty is estimated solely from the SNR. This approximation has been widely adopted in previous standard-siren forecasting studies \cite{Cai_2017,Zhang2019_ET_standard_sirens}.

The instrumental uncertainties associated with the luminosity distance $d_L$ play a central role in the final cosmological inference. The goal of this work is to incorporate a realistic description of these uncertainties, ensuring that their magnitude and statistical behavior are modeled as accurately as possible. This introduces some scatter in the simulated $d_L$ values, although the variations remain modest. What ultimately matters is that the error model captures the expected measurement precision in a more realistic way, allowing the resulting forecasts and parameter constraints to reflect the performance of the detectors rather than artifacts of an overly simplified uncertainty treatment.

The SNR-based approximation neglects parameter correlations and
includes a factor of two to approximately account for the
distance--inclination degeneracy, as discussed in more detail
in the next section. Figure~\ref{fig:luminosity_distance} shows
how the scatter in the simulated distances varies with redshift
for this method. 

In addition to instrumental noise, two astrophysical effects contribute to the total uncertainty in the luminosity distance, namely weak gravitational lensing and peculiar velocities of the host galaxy. Following common practice in the literature \cite{Zhang2019_ET_standard_sirens}, we model the fractional error due to weak lensing as
\begin{equation}
\sigma_{d_L}^{\mathrm{lens}} = 0.05\, z \, d_L,
\end{equation}
The uncertainty associated with the peculiar motion \cite{Jin_2022} of the source is given by
\begin{equation}
\sigma_{d_L}^{\mathrm{pec}} = d_L \left[ 1 + \frac{c(1+z)}{H(z)\, d_L} \right] \frac{\sigma_v}{c},
\end{equation}
where $\sigma_v$ is the one-dimensional velocity dispersion of galaxies, taken here to be $\sigma_v = 500\, \mathrm{km\, s^{-1}}$.

The total uncertainty in the luminosity distance is then computed by adding in quadrature the contributions from instrumental noise, weak lensing, and peculiar velocities
\begin{equation}
\sigma_{d_L}^{\mathrm{tot}} = \sqrt{ \left(\sigma_{d_L}^{\mathrm{inst}}\right)^2 + \left(\sigma_{d_L}^{\mathrm{lens}}\right)^2 + \left(\sigma_{d_L}^{\mathrm{pec}}\right)^2 }.
\end{equation}

\section{Methodology}
\label{sec:methodology}

The estimation of gravitational-wave parameter uncertainties was performed using the \textit{GW Doublet Approximation} implemented in GWDALI \cite{de_Souza_2023}. This framework provides an efficient approach to forecast the precision with which intrinsic and extrinsic parameters of gravitational-wave sources can be measured. GWDALI is based on the Derivative Approximation for Likelihoods (DALI), which extends the standard Fisher-matrix formalism by incorporating higher-order information from waveform derivatives, at a substantially lower computational cost than a full parameter estimation analysis.

The standard Fisher-matrix approach provides a convenient starting point for estimating waveform parameter uncertainties. Considering a set of source parameters $\vec{\theta}$ and expanding the log-likelihood around its maximum, the lowest order non vanishing contribution is quadratic in the parameter deviations $\Delta\theta_i$. The likelihood can therefore be approximated as a multivariate Gaussian,
\begin{equation}
\ln \mathcal{L}(\vec{\theta}) \simeq \ln \mathcal{L}(\vec{\theta}_0)-\frac{1}{2}F_{ij}\,\Delta\theta_i\Delta\theta_j,
\end{equation}
where $\vec{\theta}_0$ denotes the fiducial parameter values, $\Delta\theta_i=\theta_i-\theta_{0,i}$, and $F_{ij}$ is the Fisher information matrix. For gravitational-wave signals with stationary Gaussian noise, its elements can be evaluated from the noise-weighted inner products of first waveform derivatives,
\begin{equation}
F_{ij}=\left\langle\frac{\partial h}{\partial\theta_i}\bigg|\frac{\partial h}{\partial\theta_j}\right\rangle.
\end{equation}
The Fisher matrix characterizes the local sensitivity of the waveform to small changes in the source parameters. Under the Gaussian approximation and assuming an invertible Fisher matrix without additional informative priors, the parameter covariance matrix is approximated by
\begin{equation}
C_{ij}=(F^{-1})_{ij},
\end{equation}
and the marginalized $1\sigma$ uncertainty associated with a parameter $\theta_i$ is estimated as
\begin{equation}
\sigma_{\theta_i}=\sqrt{(F^{-1})_{ii}}.
\end{equation}

The inversion of the full Fisher matrix accounts for correlations between waveform parameters within the Gaussian approximation. If these correlations are neglected, the uncertainty can instead be approximated from the corresponding diagonal Fisher element. For the luminosity distance, whose waveform amplitude scales as $h\propto d_L^{-1}$ when the remaining parameters are fixed, this gives
\begin{equation}
\sigma_{d_L}^{\mathrm{inst}}\simeq\frac{1}{\sqrt{F_{d_Ld_L}}}=\frac{d_L}{\sqrt{\langle h|h\rangle}}=\frac{d_L}{\rho_{\mathrm{net}}}.
\end{equation}
This is an idealized estimate that neglects the correlations of $d_L$ with the other source parameters. A frequently used forecasting prescription introduces an additional factor of two to approximately account for the distance--inclination degeneracy, yielding $\sigma_{d_L}^{\mathrm{inst}}\simeq 2d_L/\rho_{\mathrm{net}}$ \cite{Cai_2017,Zhang2019_ET_standard_sirens}. We refer to this expression as the SNR-based method and use it only for comparison with the GWDALI distance uncertainties, as introduced in Eq.~\eqref{eq:dl_error}.

Although computationally efficient, the Fisher approximation can become unreliable when the likelihood departs significantly from a multivariate Gaussian or when strong parameter degeneracies produce nearly flat directions. This is particularly relevant for the luminosity distance $d_L$ and inclination angle $\iota$, whose correlation can lead to a poorly conditioned or nearly singular Fisher matrix for binaries approaching a face-on configuration \cite{de_Souza_2023}. In such cases, direct inversion may be numerically unstable and the Gaussian approximation may not adequately describe the marginalized uncertainties.

The DALI formalism extends this approach by retaining higher-order information in the expansion of the waveform. In the Doublet-DALI approximation adopted here, second derivatives of the waveform are included in addition to the first derivatives entering the Fisher matrix. The resulting approximation contains quadratic, cubic, and quartic terms in the parameter deviations, allowing departures from Gaussianity to be characterized beyond the Fisher approximation \cite{de_Souza_2023}. The corresponding log-likelihood can be written as
\begin{equation}
\begin{aligned}
\log \mathcal{L}_{\mathrm{Doublet}}\simeq\,&\log \mathcal{L}_0
-\frac{1}{2}\left\langle\partial_a h\mid\partial_b h\right\rangle\Delta\theta^a\Delta\theta^b\\
&-\frac{1}{2}\left\langle\partial_a h\mid\partial_b\partial_c h\right\rangle\Delta\theta^a\Delta\theta^b\Delta\theta^c\\
&-\frac{1}{8}\left\langle\partial_a\partial_b h\mid\partial_c\partial_d h\right\rangle\Delta\theta^a\Delta\theta^b\Delta\theta^c\Delta\theta^d,
\end{aligned}
\end{equation}
where $\Delta\theta^a=\theta^a-\theta_0^a$ denotes the displacement from the fiducial parameter values. The first term beyond $\log\mathcal{L}_0$ corresponds to the Fisher approximation, whereas the remaining terms provide the higher-order Doublet-DALI contributions.

The parameter uncertainties are then determined from the marginalized distributions constructed within the Doublet-DALI approximation rather than from the inverse Fisher matrix. In this work, they are characterized using the central $68\%$ credible intervals. The resulting luminosity-distance uncertainty defines the instrumental contribution $\sigma_{d_L}^{\mathrm{inst}}$ used in the cosmological forecasts.

The waveform model is described by the parameter vector $\vec{\theta}=(\alpha,\delta,{\cal M},q,\iota,d_L,\psi,\phi_c,t_c,a_1,a_2)$, where $\alpha$ is the right ascension, $\delta$ the declination, ${\cal M}$ the chirp mass, $q$ the mass ratio, $\iota$ the inclination angle, $\psi$ the polarization angle, $\phi_c$ the phase at coalescence, $t_c$ the coalescence time, and $a_1$ and $a_2$ the dimensionless aligned-spin parameters of the two compact objects. This vector lists the waveform parameters; not all of them are varied in the uncertainty estimation described below. All angles are expressed in radians and masses in solar masses. The source-frame chirp mass is defined as
\begin{equation}
{\cal M}=\frac{(m_1m_2)^{3/5}}{(m_1+m_2)^{1/5}}.
\label{eq:chirp_mass}
\end{equation}
The chirp mass entering the observed gravitational-wave signal is the detector-frame quantity ${\cal M}_z=(1+z){\cal M}$. For bright standard sirens, the electromagnetic counterpart provides an independent measurement of the redshift, permitting the source-frame chirp mass to be inferred from the detector-frame value.

We consider non-spinning binary injections and keep both the coalescence time $t_c$, expressed in GPS seconds, and the phase at coalescence $\phi_c$ fixed in the uncertainty analysis, thereby reducing the dimensionality of the parameter space.
Our analysis is divided into two parts. First, we investigate how the maximum redshift, $z_{\max}$, and maximum inclination, $\iota_{\max}$, affect the resulting cosmological constraints. We then examine the effect of the binary mass ratio, $q$, under fixed redshift and inclination limits.

In the first part, the component masses are sampled within $m_{1,2}\in[1.0,2.2]\,M_\odot$, with no explicit restriction imposed on the mass ratio. Consequently, $q$ varies from event to event according to the sampled component masses and is not used as an independent parameter to define the catalogues in this part of the analysis. We adopt the convention $q=m_1/m_2\leq1$, with $m_1\leq m_2$.

We define six configurations corresponding to all combinations of $z_{\max}\in\{0.5,1.0,2.0\}$ and $\iota_{\max}\in\{20^\circ,40^\circ\}$. For each configuration, the catalogue size $N$ is varied according to two sequences: from $N=40$ to $N=400$ in increments of $\Delta N=40$, and from $N=200$ to $N=1000$ in increments of $\Delta N=200$. The overlapping values $N=200$ and $N=400$ are included only once. This sampling provides finer resolution for smaller catalogues while extending the analysis to populations containing up to 1000 detected bright standard sirens.

This construction enables two complementary comparisons. At fixed $z_{\max}$ and $\iota_{\max}$, varying $N$ shows how the cosmological precision improves as information from additional events is accumulated. Conversely, comparing different redshift and inclination limits at fixed $N$ isolates their effect on the cosmological constraining power of the selected population. Since the number of sources is held fixed in this comparison, differences between configurations arise from the redshift and inclination distributions of the selected binaries rather than from differences in the total number of detected events.

The second part isolates the effect of the binary mass ratio by considering three representative values, $q=0.455$, $q=0.636$, and $q=0.818$, while fixing the catalogue selection at $z_{\max}=1.0$ and $\iota_{\max}=20^\circ$. For each value of $q$, we use the same catalogue size sequences adopted in the first part. This allows us to examine both the evolution of cosmological precision with $N$ and the effect of $q$ at fixed catalogue size, independently of changes in the redshift and inclination limits.

For each simulated catalogue, the individual luminosity-distance measurements are combined with the corresponding source redshifts to constrain the cosmological parameters. It is important to distinguish between parameters whose injected values vary across events and those treated as free parameters in the GWDALI analysis. In particular, the right ascension $\alpha$ and declination $\delta$ vary when generating the source population, ensuring that the events occupy different sky positions, but are held fixed when estimating parameter uncertainties. For bright standard sirens, this assumption is motivated by the identification of an electromagnetic counterpart, which can provide accurate source localization. The luminosity distance and inclination, on the other hand, remain free parameters in the GWDALI analysis.

The parameters adopted in the simulations and their status in the GWDALI analysis are summarized in Table~\ref{tab:gw_parameters}.

\begin{table}[!htbp]
\centering
\renewcommand{\arraystretch}{1.15}
\setlength{\tabcolsep}{9pt}
\begin{tabular}{@{}lccc@{}}
\hline
\textbf{Parameter} & \textbf{Values or range} & \textbf{Units} & \textbf{Free} \\
\hline
$m_{1,2}$ & $[1.0,\,2.2]$ & $M_\odot$ & Yes \\
$q$ & $\{0.455,\,0.636,\,0.818\}$ & -- & Yes \\
$d_L$ & -- & Mpc & Yes \\
$\iota$ & $[0,\,\iota_{\max}]$ & rad & Yes \\
$\delta$ & $[-\pi/2,\,\pi/2]$ & rad & No \\
$\alpha$ & $[0,\,2\pi]$ & rad & No \\
$\psi$ & $[0,\,\pi]$ & rad & Yes \\
\hline
\end{tabular}
\caption{Parameters adopted in the simulated gravitational-wave catalogues. The last column indicates whether each parameter is treated as free in the GWDALI uncertainty analysis. Component masses in $[1.0,2.2]M_\odot$ are sampled in the first analysis, whereas the injected mass ratio is fixed to each of the three listed values in the second analysis. The luminosity distance $d_L$ is set by the redshift of each source in the fiducial cosmology. Although $\alpha$ and $\delta$ vary across events, they are held fixed during the uncertainty estimation.}
\label{tab:gw_parameters}
\end{table}

\subsection{Tidal Effects and the Waveform Approximation}
\label{sec:tidal_effects}

The original IMRPhenomD model was developed for binary black hole systems and therefore does not account for the tidal effects associated with the internal structure of neutron stars. For binary neutron star (BNS) systems, these effects become increasingly important during the late inspiral and can be incorporated through extensions such as IMRPhenomD\_NRTidal \cite{Dietrich_2017}. In the main analysis of this work, IMRPhenomD is nevertheless adopted as a simplified waveform model for the BNS population. Since the cosmological information is primarily obtained from the luminosity distance, we assess below the extent to which neglecting the tidal sector can affect the quantities relevant to our forecasts.

The response of a neutron star to the tidal field generated by its companion is characterized by the dimensionless tidal deformability $\Lambda_i$, which is related to the quadrupolar tidal Love number $k_2$ through

\begin{equation}
    \Lambda_i =
    \frac{2}{3} k_2
    \left(
        \frac{R_i c^2}{Gm_i}
    \right)^5,
    \label{eq:tidal_deformability}
\end{equation}
where $m_i$ and $R_i$ are the mass and radius of each star, respectively \cite{Hinderer_2008}. For a binary system, the leading tidal contribution is characterized by the effective tidal deformability,

\begin{equation}
    \tilde{\Lambda}
    =
    \frac{16}{13}
    \frac{
        (12q+1)\Lambda_1
        +
        (12+q)q^4\Lambda_2
    }{
        (1+q)^5
    },
    \label{eq:effective_tidal_deformability}
\end{equation}
with $q=m_1/m_2\leq1$. For neutron stars described by the same equation of state, the individual tidal deformabilities can be approximately related through

\begin{equation}
    \frac{\Lambda_2}{\Lambda_1}
    \simeq q^6,
    \label{eq:lambda_mass_relation}
\end{equation}
highlighting the connection between the mass ratio and the tidal sector \cite{De_2018}.

As a first diagnostic, we compare the SNR obtained with IMRPhenomD with that obtained using IMRPhenomD\_NRTidal across the binary parameter space. For equal mass binaries, we take $m_1=m_2=m$ and $\Lambda_1=\Lambda_2=\Lambda$. For unequal mass systems, we fix $m_1=1\,M_\odot$, vary the mass ratio, and determine $\Lambda_2$ from Eq.~\eqref{eq:lambda_mass_relation}. For this test, the tidal deformability is deliberately extended up to $\Lambda=5000$, well beyond the range typically favored by current BNS observations, in order to probe extreme tidal configurations.

\begin{figure}[t]
    \centering
    \includegraphics[width=\linewidth]{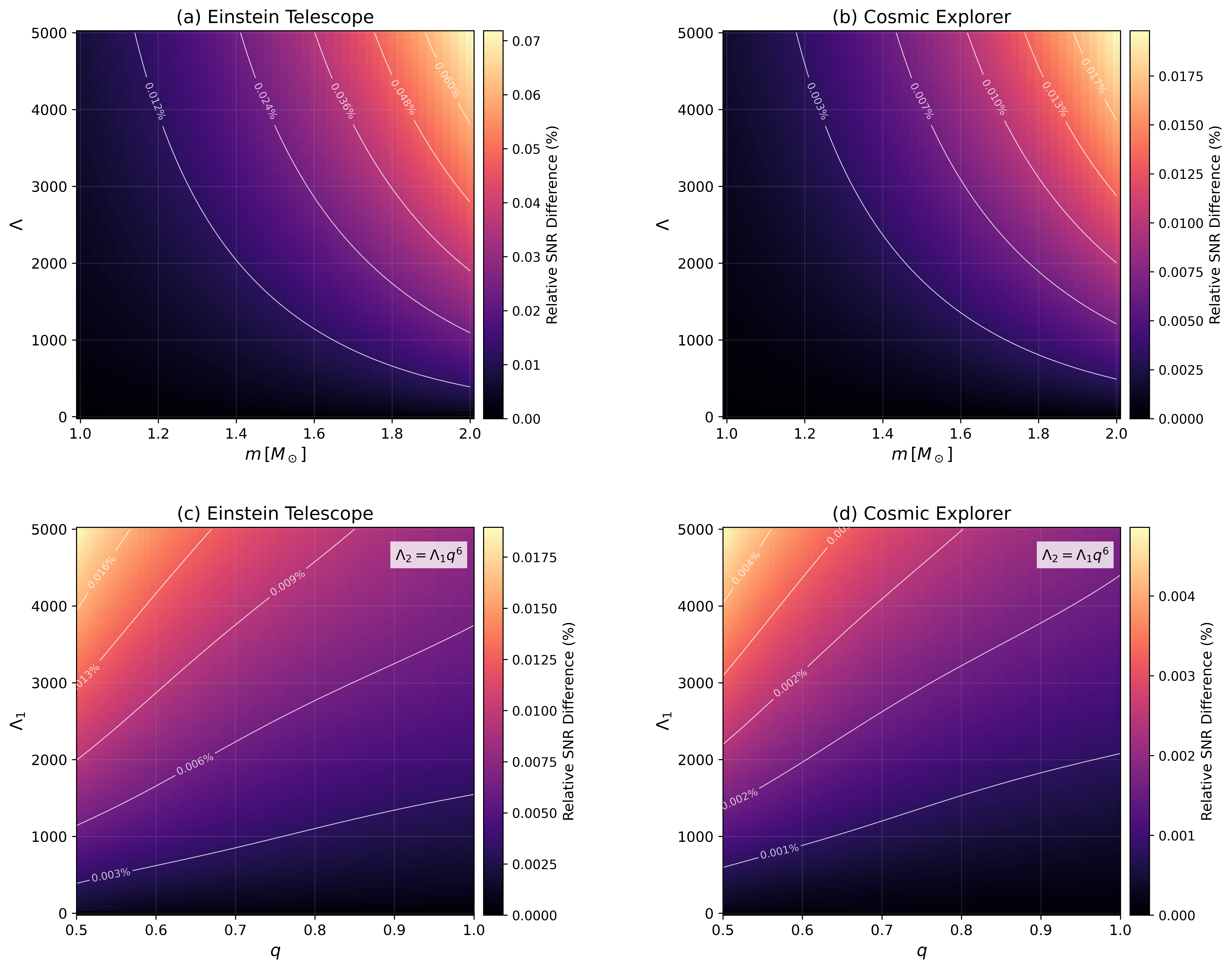}
    \caption{Relative difference between the SNR obtained with IMRPhenomD and IMRPhenomD\_NRTidal across the binary parameter space. Panels (a) and (b) correspond to equal mass binaries, $m_1=m_2=m$ and $\Lambda_1=\Lambda_2=\Lambda$, for the Einstein Telescope and Cosmic Explorer, respectively. Panels (c) and (d) correspond to unequal mass binaries with $m_1=1\,M_\odot$ and $q\leq1$, where $\Lambda_1$ is varied and $\Lambda_2$ is determined according to Eq.~\eqref{eq:lambda_mass_relation}.}
    \label{fig:relative_snr_parameter_space}
\end{figure}

As shown in Fig.~\ref{fig:relative_snr_parameter_space}, the inclusion of tidal corrections produces only a very small change in the SNR over the explored parameter space. For the Einstein Telescope, the relative difference remains below approximately $0.1\%$, while for the Cosmic Explorer it remains below approximately $0.02\%$, even for the deliberately large tidal deformabilities considered here.

If the impact of the waveform approximation were assessed exclusively through the SNR, these results could therefore suggest that the difference between the tidal and nontidal waveform descriptions is negligible for the present forecasts. However, the SNR measures the overall strength of the signal, whereas the luminosity-distance uncertainty obtained from a multidimensional inference also depends on the structure of the likelihood and on correlations with the remaining parameters. A small change in the total SNR therefore does not necessarily imply an equally small change in the marginalized uncertainty on $D_L$, particularly when $\Lambda_1$ and $\Lambda_2$ are introduced as additional inference parameters.

\begin{figure}[t]
    \centering
    \includegraphics[width=0.80\linewidth]{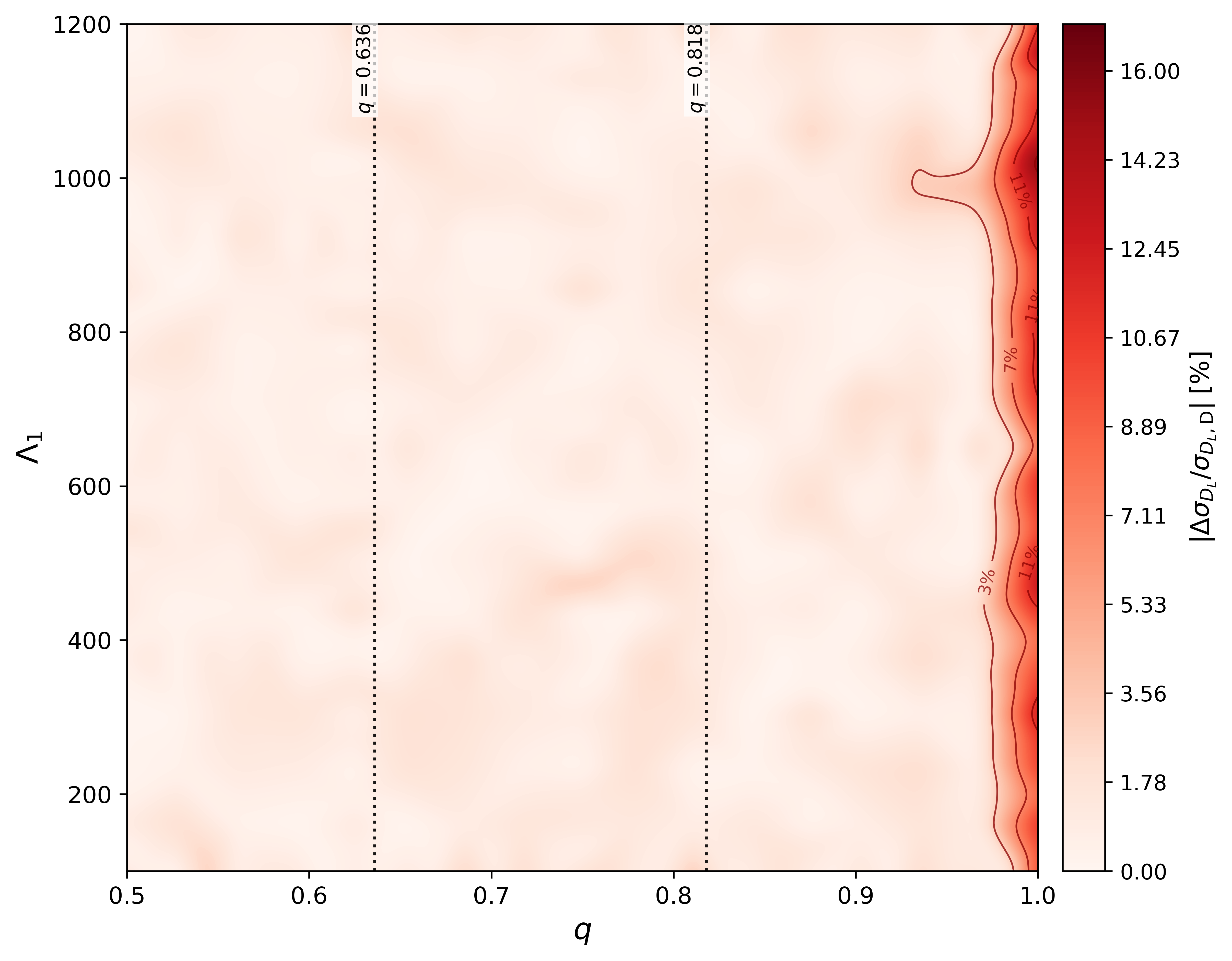}
    \caption{Absolute relative difference in the marginalized luminosity-distance uncertainty obtained from the tidal and nontidal waveform analyses as a function of $q$ and $\Lambda_1$. The contours indicate differences of $3\%$, $7\%$, $11\%$, and $15\%$. The vertical dotted lines indicate the mass ratio values of the main analysis that lie within the range covered by the tidal test.}
    \label{fig:tidal_distance_map}
\end{figure}

To investigate this effect, we repeat the analysis allowing the individual tidal deformabilities to vary and compare the resulting luminosity-distance uncertainties. For this test, we consider the more physically relevant interval $100\leq\Lambda_1\leq1200$ and $0.5\leq q\leq1$, with $m_1=1\,M_\odot$, $m_2=m_1/q$, and assuming the relation given by Eq.~\ref{eq:lambda_mass_relation}. For each point in the $(q,\Lambda_1)$ grid, we perform the inference using both IMRPhenomD and IMRPhenomD\_NRTidal and determine the corresponding marginalized uncertainty on $D_L$. These calculations are performed using the Triplet DALI approximation implemented in GWDALI, which includes higher order terms in the likelihood expansion and provides a more accurate characterization of the posterior than the lower order approximation used in the main analysis. In the tidal analysis, $\Lambda_1$ and $\Lambda_2$ are included as additional inference parameters, whereas the nontidal analysis is performed without these degrees of freedom. We then compute the absolute relative difference at each grid point. The resulting map is shown in Fig.~\ref{fig:tidal_distance_map}. In contrast with the SNR comparison, the effect is no longer uniformly negligible. Across most of the mass ratio range, the difference remains small, with only statistical variations across the explored values of $q$ and $\Lambda_1$. A much stronger and systematic increase develops as the system approaches the equal mass limit, $q\rightarrow1$, where the difference reaches values above $10\%$ for some tidal configurations.

\begin{figure}[t]
    \centering
    \includegraphics[width=0.95\linewidth]{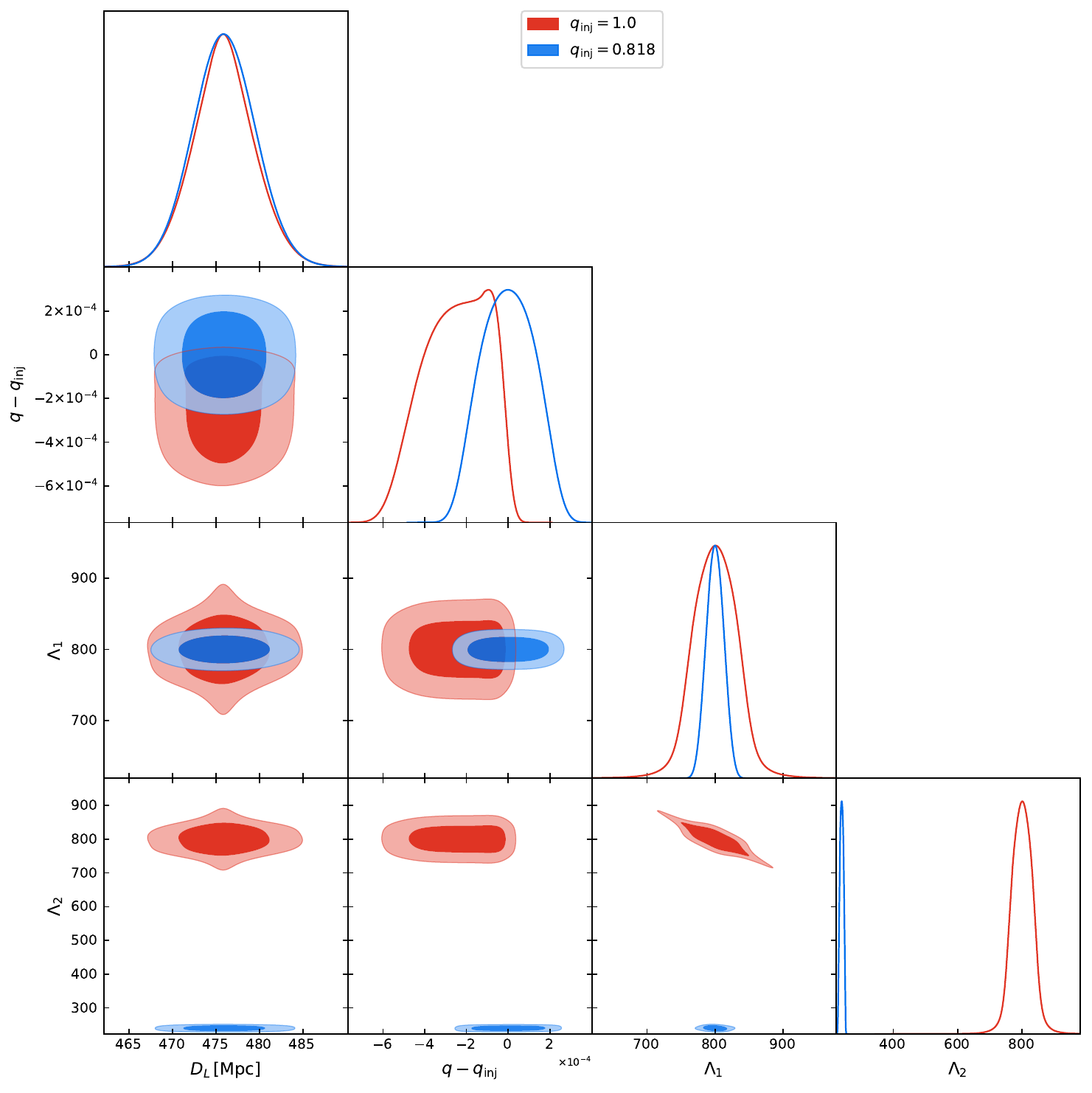}
    \caption{Joint marginalized posterior distributions for $q_{\rm inj}=1$ and $q_{\rm inj}=0.818$ when the individual tidal deformabilities are included in the inference. The mass ratio coordinate is displayed as $\Delta q=q-q_{\rm inj}$. The equal mass configuration exhibits a strong anticorrelation between $\Lambda_1$ and $\Lambda_2$, which is substantially reduced for $q_{\rm inj}=0.818$.}
    \label{fig:tidal_triangle}
\end{figure}

The localization of this effect close to $q=1$ can be understood from the corresponding joint posterior distributions. Figure~\ref{fig:tidal_triangle} compares an equal mass injection, $q_{\rm inj}=1$, with the largest mass ratio explicitly considered in our mass ratio study, $q_{\rm inj}=0.818$. For visualization, the mass ratio coordinate is shifted according to $\Delta q=q-q_{\rm inj}$, allowing the two posterior widths and correlations to be displayed on a common scale. In the equal mass case, the posterior is truncated at $\Delta q=0$ by the physical condition $q\leq1$, while a pronounced anticorrelation appears between $\Lambda_1$ and $\Lambda_2$. Near equal masses, the waveform is more sensitive to combinations of the tidal deformabilities than to the two individual parameters separately, allowing changes in $\Lambda_1$ and $\Lambda_2$ to partially compensate each other. For $q_{\rm inj}=0.818$, this symmetry is broken and the two tidal parameters become considerably more localized.

The triangle plot does not indicate a comparably strong direct correlation between $D_L$ and either individual tidal deformability. Instead, the increase in $\sigma_{D_L}$ occurs when the inference is extended to a tidal parameter space that becomes strongly degenerate near the equal mass limit. Marginalizing over these additional, poorly constrained degrees of freedom can consequently broaden the luminosity-distance posterior even though the direct modification of the total SNR is very small. 

This behavior is also important when assessing its relevance for the population used in our forecasts. The two component masses are drawn independently from $m_1,m_2\sim\mathcal{U}(1,2.2)\,M_\odot$, after which they are ordered such that the mass ratio satisfies $q\leq1$. Although the component masses are sampled uniformly, the resulting distribution of $q$ is not uniform and increases toward the equal mass limit. To quantify both the occurrence of systems in this region and their expected contribution to the variation in the luminosity-distance uncertainty, we generate $10^6$ binaries following the same mass distribution adopted in our catalogs.

\begin{figure*}[t]
    \centering
    \includegraphics[width=\textwidth]{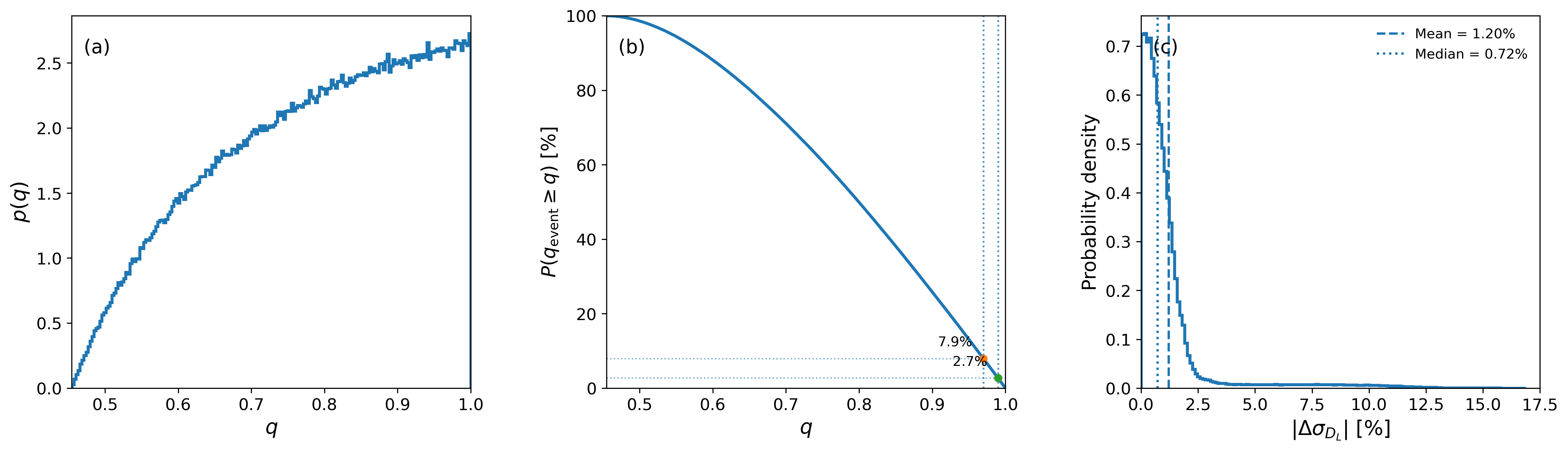}
    \caption{Population assessment of the tidal contribution to the luminosity-distance uncertainty. Panel (a) shows the mass ratio probability density resulting from independently drawing both component masses from $\mathcal{U}(1,2.2)\,M_\odot$. Panel (b) shows the corresponding upper tail cumulative probability $P(q_{\rm event}\geq q)$, with $q=0.97$ and $q=0.99$ highlighted to indicate the fraction of systems approaching the equal mass limit. Panel (c) shows the distribution of the absolute relative change in the luminosity-distance uncertainty, $|\Delta\sigma_{D_L}|$, obtained by additionally sampling $\Lambda_1\sim\mathcal{U}(100,1200)$ and interpolating the tidal grid of Fig.~\ref{fig:tidal_distance_map}. The vertical lines indicate the mean and median of the resulting distribution.}
    \label{fig:q_population}
\end{figure*}

The resulting population analysis is shown in Fig.~\ref{fig:q_population}. Although the two component masses are sampled independently from uniform distributions, their ratio does not follow a uniform distribution. As shown in panel (a), the resulting mass ratio distribution becomes increasingly concentrated toward $q=1$. Panel (b) shows the corresponding upper tail cumulative probability, $P(q_{\rm event}\geq q)$, which represents the fraction of binaries in the population with a mass ratio equal to or larger than a given value of $q$. Despite the increasing probability density toward equal masses, only approximately $7.9\%$ of the systems satisfy $q\geq0.97$, and this fraction decreases to $2.7\%$ for $q\geq0.99$. Thus, although binaries become more densely distributed toward larger $q$, systems lying extremely close to the equal mass limit still constitute only a small fraction of the total population.

To estimate the corresponding effect on the luminosity-distance uncertainty, we additionally sample $\Lambda_1$ uniformly over the interval $100\leq\Lambda_1\leq1200$ and assign to each simulated binary the value of $|\Delta\sigma_{D_L}|$ obtained by interpolating the $(q,\Lambda_1)$ grid shown in Fig.~\ref{fig:tidal_distance_map}. This provides an estimate of the expected change in the distance uncertainty for the adopted mass and tidal deformability distributions. The resulting distribution is shown in panel (c) of Fig.~\ref{fig:q_population}.

The resulting distribution of $|\Delta\sigma_{D_L}|$ is strongly concentrated at small values, with a median of approximately $0.72\%$ and a mean of approximately $1.20\%$. About $93.4\%$ of the evaluated systems exhibit changes smaller than $3\%$, while only approximately $3.3\%$ exceed $7\%$ and $1.2\%$ exceed $10\%$. The tail toward larger values is associated predominantly with systems approaching the equal mass limit. For example, for $q\geq0.98$ the mean change increases to approximately $7.8\%$, while systems in this range constitute only about $5.3\%$ of the adopted mass population. Thus, although the additional tidal degrees of freedom can produce substantial changes in the luminosity-distance uncertainty for individual systems close to $q=1$, the expected effect over the population considered here remains small.

This result is particularly relevant for the fixed mass ratio configurations investigated in our main analysis. The values $q=0.455$, $0.636$, and $0.818$ all remain well separated from the region close to $q=1$ where the degradation in $\sigma_{D_L}$ becomes significant. The effect identified above is therefore not expected to appreciably affect the conclusions obtained from these fixed $q$ configurations. For the catalogs in which the component masses are instead sampled uniformly, systems close to equal masses are naturally present, but the distribution shown in Fig.~\ref{fig:q_population} indicates that their contribution is sufficiently limited that the mean absolute change in $\sigma_{D_L}$ remains at approximately $1.2\%$.

We therefore retain IMRPhenomD as the baseline waveform for the cosmological forecasts presented here. Our tests indicate that the largest changes in the luminosity-distance uncertainty are concentrated near the equal mass limit, while the impact remains small over most of the parameter space relevant to our analysis. A fully consistent treatment of the tidal degrees of freedom would require a more detailed modeling of the neutron star properties and their equation of state. Since the primary goal of this work is to investigate the cosmological information provided by BNS standard sirens, such an extension lies beyond the scope of the present analysis and is left for future work.

\section{Results}

\begin{figure}[!htbp]
    \centering
    \includegraphics[width=0.8\linewidth]{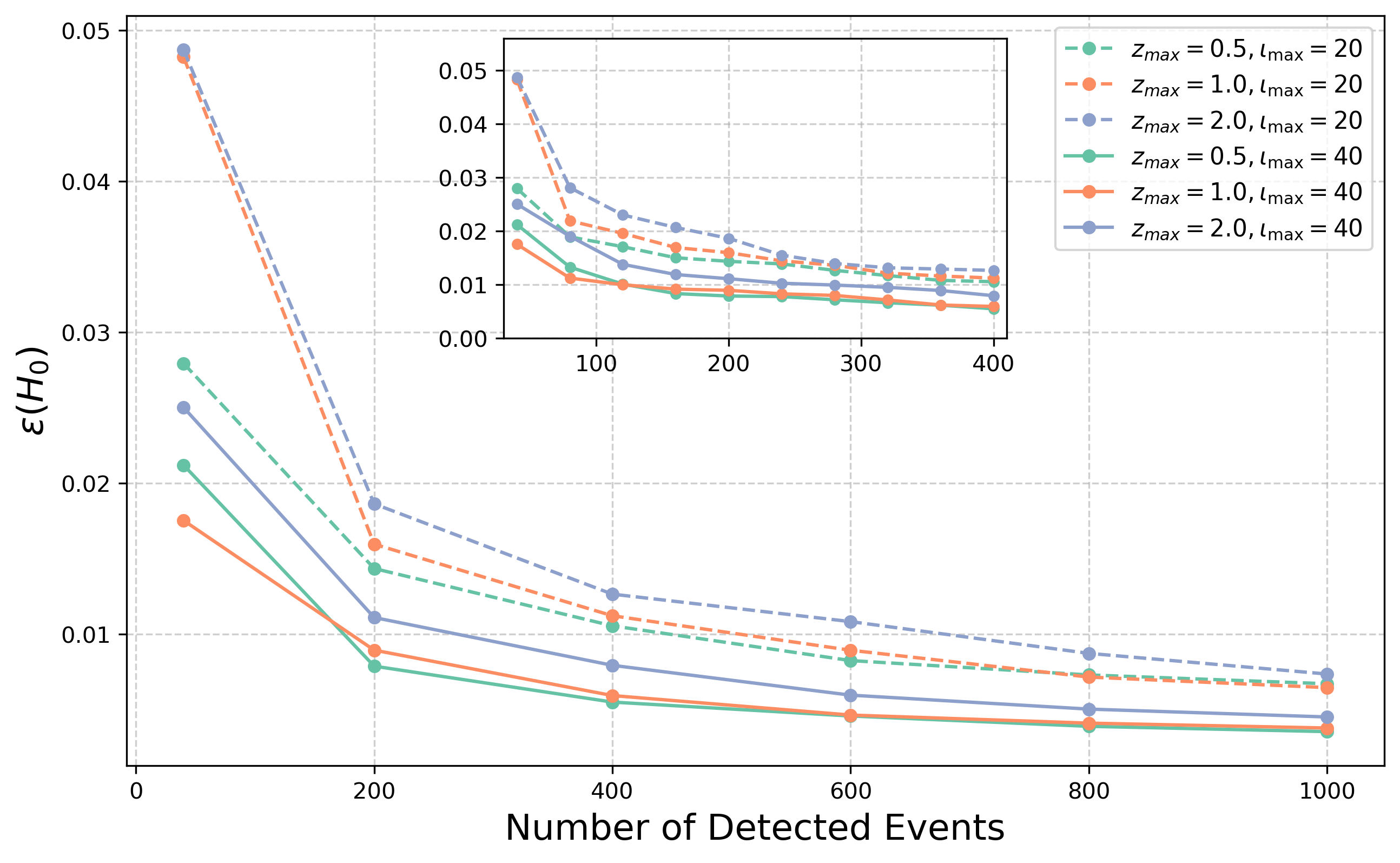}
    \caption{Relative errors of the Hubble constant $H_0$ as a function of the number of observed events $N$ for different maximum inclination angles ($\iota_\mathrm{max}$) and maximum redshifts ($z_\mathrm{max}$). The main plot shows the relative errors $\mathcal{E}(H_0)$ for selected points $N = 40, 200, 400, 600, 800, 1000$, while the inset provides a zoom of the interval $40 \leq N \leq 400$ to better visualize the behavior at lower $N$. Data for $\iota_\mathrm{max} = 20^\circ$ are displayed with dashed lines, and $\iota_\mathrm{max} = 40^\circ$ with solid lines. Colors indicate different redshifts: blue for $z_\mathrm{max} = 0.5$, green for $z_\mathrm{max} = 1$, and dark gray for $z_\mathrm{max} = 2$. Markers indicate individual simulation points, with black edges for clarity.}
    \label{fig:h0_relative_error}
\end{figure}

The results show clear and systematic tendencies in the inferred cosmological parameters as functions of both the maximum inclination $\iota_{\max}$ and the maximum redshift $z_{\max}$, as summarized in Table~\ref{tab:results_all_redshifts_iota_E}. As shown in Fig.~\ref{fig:h0_relative_error}, for a fixed number of detected events, lower-redshift samples generally provide tighter constraints on $H_0$. As $N$ approaches 1000, the statistical uncertainty becomes largely insensitive to redshift, and in the case of $z_{max}<2$, we can see this effect more clearly, with constraints on $H_0$ being slightly worse. The improvement is evident, with the relative uncertainty
on $H_0$ falling below the $1\%$ level for $N \geq 400$
and reaching $0.5\%$ or better for $N \geq 800$ when
$\iota_{\max}=40^\circ$. For $\iota_{\max}=20^\circ$,
sub-percent precision is achieved for $N \geq 800$,
regardless of the maximum redshift considered,
with relative uncertainties ranging from $0.72\%$
to $0.87\%$ at $N=800$ and from $0.65\%$
to $0.74\%$ at $N=1000$.

\begin{figure}[!htbp]
    \centering
    \includegraphics[width=0.8\linewidth]{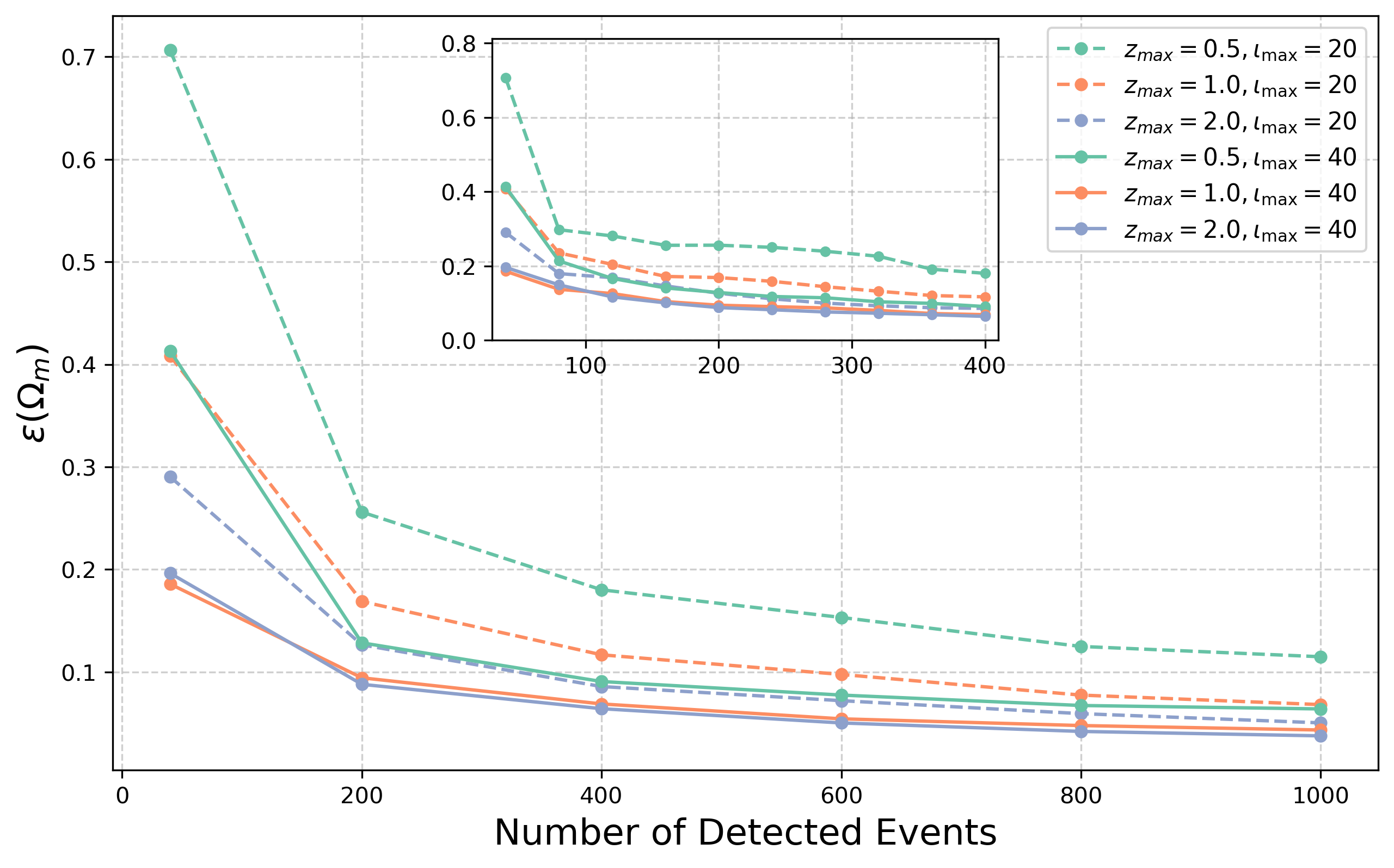}
    \caption{Relative errors of the matter density parameter $\Omega_m$ as a function of the number of observed events $N$ for different maximum inclination angles $\iota_\mathrm{max}$ and maximum redshifts $z_\mathrm{max}$. The main plot shows the relative errors for selected points $N = 40, 200, 400, 600, 800, 1000$, while the inset provides a zoom of the interval $40 \leq N \leq 400$ to better visualize the behavior at lower $N$. Data for $\iota_\mathrm{max} = 20^\circ$ are displayed with dashed lines, and $\iota_\mathrm{max} = 40^\circ$ with solid lines. Colors indicate different redshifts: blue for $z_\mathrm{max} = 0.5$, green for $z_\mathrm{max} = 1$, and dark gray for $z_\mathrm{max} = 2$. Markers indicate individual simulation points, with black edges for clarity. The relative errors are calculated as $\mathcal{E}(\Omega_m) = \sigma_{\Omega_m} /  \Omega_m$, where $\sigma_{\Omega_m}$ is the standard deviation of the estimated $\Omega_m$ values across simulations.}

    \label{fig:omg_m_relative_error}
\end{figure}

The behavior of $\Omega_m$ shows a complementary pattern (Fig.~\ref{fig:omg_m_relative_error}). Its constraints improve more gradually than those of $H_0$, reflecting the weaker sensitivity of the luminosity distance to the matter density parameter at lower redshifts. The relative uncertainty decreases as the number of detected events increases, reaching values below $10\%$ for several configurations with $N \geq 400$. For larger catalogues, the constraints continue to improve steadily, with the uncertainty further reduced by extending the redshift range and increasing the maximum detectable inclination angle. At $N=1000$, the relative uncertainty takes values between $3.92\%$ and $11.44\%$ across the six configurations considered.

Varying the inclination limit produces consistent differences across all redshifts. The simulations with $\iota_{\max}=40^\circ$ invariably yield smaller uncertainties than those with $\iota_{\max}=20^\circ$, reflecting the inclusion of more inclined systems for which the distance--inclination degeneracy is less pronounced. For example, at $z_{\max}=0.5$ and $N=1000$, $\sigma_{H_0}$ decreases from $0.456$ to $0.240\,\mathrm{km\,s^{-1}\,Mpc^{-1}}$, while the relative uncertainty on $\Omega_m$ decreases from $11.44\%$ to $6.41\%$ when moving from $\iota_{\max}=20^\circ$ to $40^\circ$. This pattern persists at higher redshifts. At $z_{\max}=2.0$, the relative uncertainty on $\Omega_m$ for $N=1000$ decreases from $5.06\%$ to $3.92\%$ under the same change in inclination range.

The variation with redshift shows a complementary behavior. Across all redshift ranges, the inferred values of $H_0$ approach the fiducial value as the number of detected events increases. For larger catalogues, the relative deviations remain small, while lower-redshift samples generally provide tighter constraints on $H_0$ for a fixed number of events.

Constraints on $\Omega_m$ strengthen substantially as $z_{\max}$ increases. This trend is expected, since although the fractional dependence of the luminosity distance on $H_0$ remains unchanged, its sensitivity to $\Omega_m$ becomes more pronounced at higher redshifts, helping to break the degeneracy between these cosmological parameters. As a result, the relative uncertainty on $\Omega_m$ decreases by approximately a factor of two when moving from $z_{\max}=0.5$ to $z_{\max}=2.0$, depending on the inclination selection and for fixed $N$. These trends indicate that extending the redshift coverage and including more inclined systems substantially improves the precision on $\Omega_m$, while $H_0$ benefits primarily from increasing the number of detected events.

\begin{figure}[!htbp]
    \centering
    \includegraphics[width=0.8\linewidth]{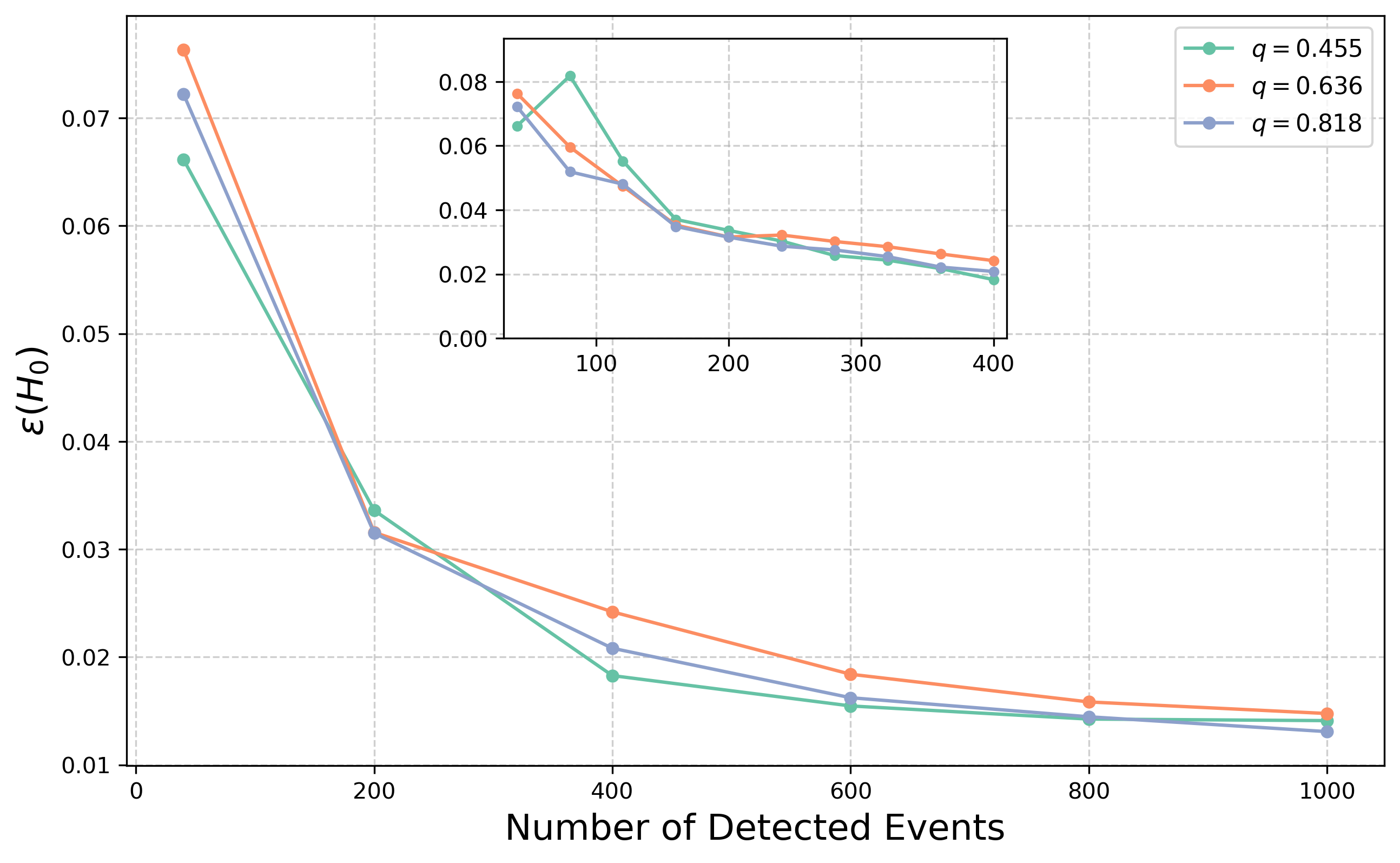}
    \caption{Relative error in the Hubble constant, ${\cal E} (H_0)$, as a function of the number of detected events $N$, for different mass ratios: $q = 0.455$, $q = 0.636$, and $q = 0.818$. The main plot shows selected points up to $N=1000$, while the inset highlights the range $40 \leq N \leq 400$.}

    \label{fig:h0_q}
\end{figure}

\begin{figure}[!htbp]
    \centering
    \includegraphics[width=0.8\linewidth]{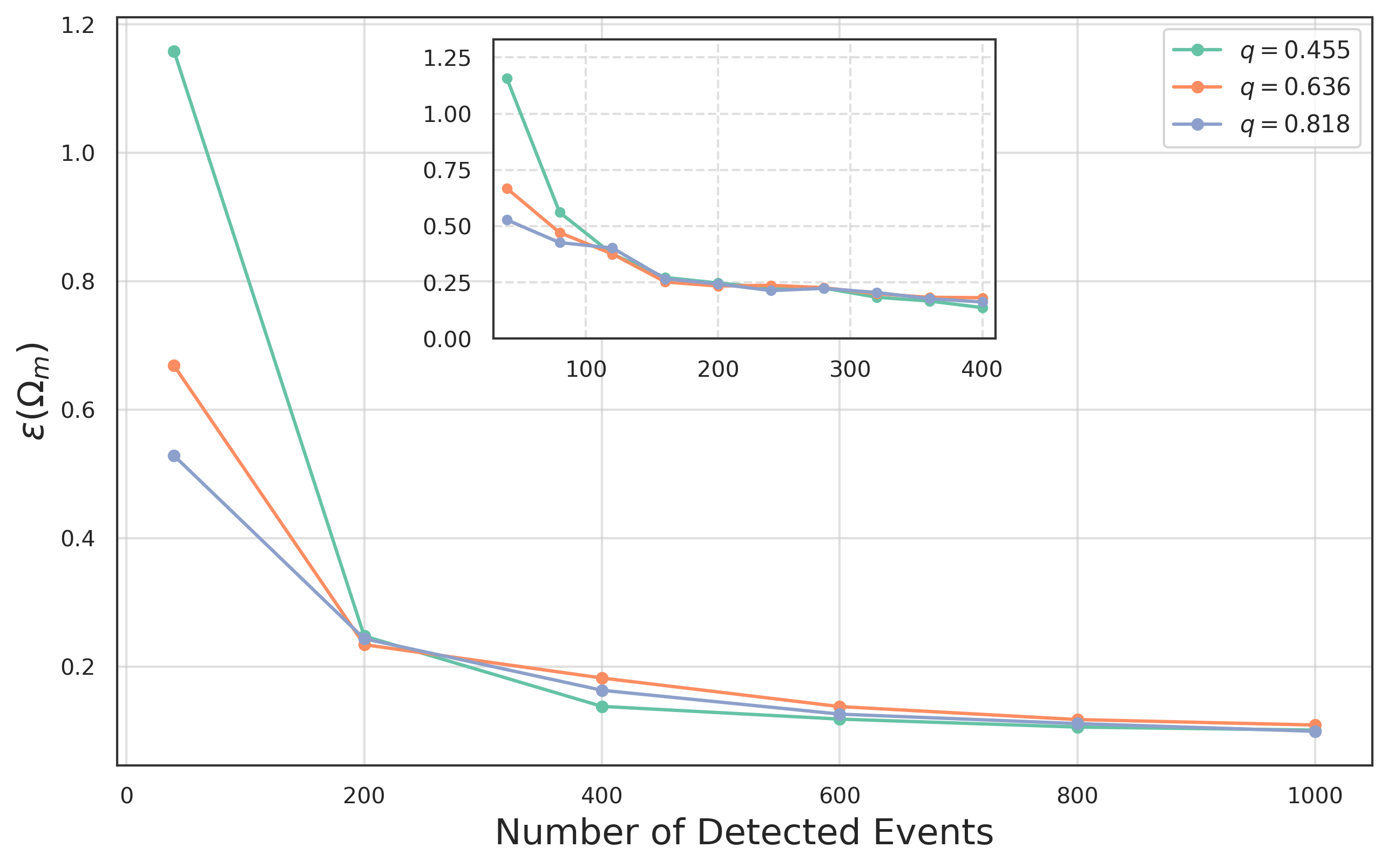}
    \caption{Relative error in the matter density parameter, ${\cal E}(\Omega_m)$, as a function of the number of detected events $N$, for different mass ratios: $q = 0.455$, $q = 0.636$, and $q = 0.818$. The main plot shows selected points up to $N=1000$, while the inset highlights the range $40 \leq N \leq 400$.}
    \label{fig:omega_m_q}
\end{figure}

Finally, we assess the impact of varying the mass ratio $q$ on the inferred cosmological constraints, as summarized in Table~\ref{tab:q} and illustrated in Figs.~\ref{fig:h0_q} and \ref{fig:omega_m_q}. At low numbers of detected events, the inferred uncertainties are dominated by statistical fluctuations, which are significantly larger than any variation induced by changing $q$. As a consequence, differences between the constraints obtained for distinct mass ratios in the low-$N$ regime should not be over-interpreted. The datasets with larger $N$ therefore provide more reliable and statistically robust estimates of the uncertainties.

For large event samples, the results obtained for different values of $q$ become broadly consistent. In particular, the constraints on $\Omega_m$ show no statistically significant dependence on the choice of mass ratio, with both $\sigma_{\Omega_m}$ and ${\cal E}(\Omega_m)$ remaining consistent across the different $q$ values and well within statistical uncertainties, as shown in Fig.~\ref{fig:omega_m_q}. This indicates that the inference of $\Omega_m$ is effectively insensitive to $q$ in the regime explored here.

For $H_0$, however, a small but perceptible difference between datasets with different mass ratios persists even at large $N$. As shown in Fig.~\ref{fig:h0_q}, both $\sigma_{H_0}$ and the relative error ${\cal E}(H_0)$ exhibit a weak non-linear dependence on the chosen value of $q$. This effect impacts the inferred uncertainty rather than the central value of $H_0$, which remains confined to a narrow range around $68\,\mathrm{km\,s^{-1}\,Mpc^{-1}}$, consistent with the Planck value of $67.66\,\mathrm{km\,s^{-1}\,Mpc^{-1}}$. Although the magnitude of this non-linearity is modest and subdominant compared to the overall statistical uncertainty, its presence suggests that the response of the $H_0$ uncertainty to variations in $q$ is not purely linear.

\begin{table}[!ht]

\scriptsize
\setlength{\tabcolsep}{4pt}
\renewcommand{\arraystretch}{0.92}
\caption{Results for the average Hubble constant $H_0 $ (in km s$^{-1}$ Mpc$^{-1}$) and average matter density parameter $ \Omega_m $ for different numbers of events $N$. 
The table reports the average errors $\sigma_{H_0}$ and $\sigma_{\Omega_m}$, and the relative errors $\mathcal{E}(H_0)$ and $\mathcal{E}(\Omega_m)$. 
Values are given for three maximum redshifts: $z=0.5$, $z=1.0$, and $z=2.0$, and for two maximum inclinations: $\iota=20^\circ$ and $\iota=40^\circ$. 
Each row corresponds to a specific number of events $N$ used in the simulation.}
\label{tab:results_all_redshifts_iota_E}
\setlength{\tabcolsep}{5.5pt}
\renewcommand{\arraystretch}{0.95}

\begin{tabular}{|c|c|cccccc|cccccc|}
\hline
\multirow{2}{*}{$z$} & \multirow{2}{*}{$N$} 
& \multicolumn{6}{c|}{$\iota=20^\circ$} 
& \multicolumn{6}{c|}{$\iota=40^\circ$} \\
 & & $H_0$ & $\sigma_{H_0}$ & ${\cal E}(H_0)$ & $\Omega_m$ & $\sigma_{\Omega_m}$ & ${\cal E} (\Omega_m)$
   & $H_0$ & $\sigma_{H_0}$ & ${\cal E}(H_0)$ & $\Omega_m$ & $\sigma_{\Omega_m}$ & ${\cal E} (\Omega_m)$ \\
\hline
\multirow{13}{*}{0.5}
 & 40   & 68.396 & 1.911 & 2.79\% & 0.182 & 0.129 & 70.88\% & 69.071 & 1.464 & 2.12\% & 0.222 & 0.092 & 41.44\% \\
 & 80   & 66.093 & 1.250 & 1.89\% & 0.374 & 0.111 & 29.68\% & 67.036 & 0.890 & 1.33\% & 0.345 & 0.074 & 21.45\% \\
 & 120  & 67.569 & 1.154 & 1.71\% & 0.337 & 0.095 & 28.19\% & 66.843 & 0.675 & 1.01\% & 0.361 & 0.060 & 16.63\% \\
 & 160  & 68.059 & 1.022 & 1.50\% & 0.299 & 0.077 & 25.75\% & 66.804 & 0.557 & 0.83\% & 0.360 & 0.051 & 14.17\% \\
 & 200  & 68.451 & 0.983 & 1.44\% & 0.278 & 0.071 & 25.54\% & 66.889 & 0.528 & 0.79\% & 0.363 & 0.047 & 12.95\% \\
 & 240  & 68.629 & 0.952 & 1.39\% & 0.268 & 0.067 & 24.93\% & 66.715 & 0.519 & 0.78\% & 0.390 & 0.046 & 11.79\% \\
 & 280  & 68.821 & 0.872 & 1.27\% & 0.246 & 0.059 & 23.98\% & 66.907 & 0.481 & 0.72\% & 0.376 & 0.043 & 11.47\% \\
 & 320  & 68.727 & 0.804 & 1.17\% & 0.259 & 0.058 & 22.39\% & 66.941 & 0.444 & 0.66\% & 0.376 & 0.039 & 10.37\% \\
 & 360  & 68.518 & 0.740 & 1.08\% & 0.279 & 0.053 & 18.99\% & 67.061 & 0.414 & 0.62\% & 0.365 & 0.036 & 9.86\% \\
 & 400  & 68.190 & 0.720 & 1.06\% & 0.295 & 0.053 & 17.97\% & 67.175 & 0.370 & 0.55\% & 0.354 & 0.032 & 9.04\% \\
 & 600  & 68.586 & 0.566 & 0.83\% & 0.266 & 0.041 & 15.41\% & 67.332 & 0.309 & 0.46\% & 0.335 & 0.026 & 7.76\% \\
 & 800  & 67.997 & 0.497 & 0.73\% & 0.297 & 0.037 & 12.46\% & 67.480 & 0.263 & 0.39\% & 0.327 & 0.022 & 6.73\% \\
 &1000  & 67.869 & 0.456 & 0.67\% & 0.306 & 0.035 & 11.44\% & 67.615 & 0.240 & 0.36\% & 0.312 & 0.020 & 6.41\% \\
\hline
\multirow{13}{*}{1.0}
 & 40   & 65.724 & 3.169 & 4.82\% & 0.394 & 0.161 & 40.86\% & 67.090 & 1.176 & 1.75\% & 0.334 & 0.062 & 18.56\% \\
 & 80   & 68.375 & 1.497 & 2.19\% & 0.280 & 0.066 & 23.57\% & 67.081 & 0.753 & 1.12\% & 0.327 & 0.045 & 13.76\% \\
 & 120  & 68.466 & 1.339 & 1.96\% & 0.286 & 0.058 & 20.28\% & 67.425 & 0.673 & 1.00\% & 0.289 & 0.036 & 12.46\% \\
 & 160  & 68.501 & 1.160 & 1.69\% & 0.288 & 0.050 & 17.36\% & 67.068 & 0.615 & 0.92\% & 0.315 & 0.033 & 10.48\% \\
 & 200  & 68.611 & 1.095 & 1.60\% & 0.281 & 0.047 & 16.73\% & 67.086 & 0.600 & 0.89\% & 0.318 & 0.030 & 9.43\% \\
 & 240  & 68.928 & 0.995 & 1.44\% & 0.270 & 0.043 & 15.93\% & 67.282 & 0.558 & 0.83\% & 0.310 & 0.028 & 9.09\% \\
 & 280  & 68.707 & 0.936 & 1.36\% & 0.276 & 0.040 & 14.49\% & 67.454 & 0.539 & 0.80\% & 0.302 & 0.026 & 8.61\% \\
 & 320  & 68.703 & 0.835 & 1.22\% & 0.270 & 0.036 & 13.33\% & 67.656 & 0.484 & 0.72\% & 0.298 & 0.024 & 8.05\% \\
 & 360  & 68.486 & 0.795 & 1.16\% & 0.280 & 0.034 & 12.26\% & 67.770 & 0.421 & 0.62\% & 0.294 & 0.021 & 7.14\% \\
 & 400  & 68.285 & 0.766 & 1.12\% & 0.289 & 0.034 & 11.74\% & 67.587 & 0.401 & 0.59\% & 0.302 & 0.021 & 6.95\% \\
 & 600  & 68.541 & 0.613 & 0.89\% & 0.275 & 0.027 & 9.88\% & 67.592 & 0.314 & 0.46\% & 0.309 & 0.017 & 5.50\% \\
 & 800  & 68.090 & 0.488 & 0.72\% & 0.290 & 0.023 & 7.93\% & 67.613 & 0.278 & 0.41\% & 0.304 & 0.015 & 4.93\% \\
 &1000  & 67.985 & 0.440 & 0.65\% & 0.300 & 0.020 & 6.67\% & 67.630 & 0.256 & 0.38\% & 0.311 & 0.014 & 4.50\% \\
\hline
\multirow{13}{*}{2.0}
 & 40   & 65.827 & 3.206 & 4.87\% & 0.355 & 0.103 & 29.01\% & 66.569 & 1.665 & 2.50\% & 0.313 & 0.061 & 19.49\% \\
 & 80   & 66.888 & 1.878 & 2.81\% & 0.329 & 0.059 & 17.92\% & 66.659 & 1.268 & 1.90\% & 0.310 & 0.046 & 14.84\% \\
 & 120  & 68.996 & 1.587 & 2.30\% & 0.258 & 0.043 & 16.67\% & 67.473 & 0.930 & 1.38\% & 0.292 & 0.034 & 11.64\% \\
 & 160  & 68.536 & 1.416 & 2.07\% & 0.273 & 0.040 & 14.65\% & 67.663 & 0.804 & 1.19\% & 0.297 & 0.030 & 10.10\% \\
 & 200  & 68.051 & 1.268 & 1.86\% & 0.284 & 0.036 & 12.68\% & 67.200 & 0.746 & 1.11\% & 0.319 & 0.028 & 8.77\% \\
 & 240  & 68.089 & 1.056 & 1.55\% & 0.283 & 0.032 & 11.31\% & 67.237 & 0.690 & 1.03\% & 0.324 & 0.027 & 8.33\% \\
 & 280  & 67.644 & 0.944 & 1.40\% & 0.295 & 0.029 & 9.83\% & 67.090 & 0.666 & 0.99\% & 0.329 & 0.025 & 7.60\% \\
 & 320  & 67.167 & 0.883 & 1.31\% & 0.303 & 0.028 & 9.24\% & 67.276 & 0.639 & 0.95\% & 0.328 & 0.024 & 7.32\% \\
 & 360  & 67.182 & 0.868 & 1.29\% & 0.313 & 0.027 & 8.63\% & 67.405 & 0.601 & 0.89\% & 0.319 & 0.022 & 6.90\% \\
 & 400  & 66.996 & 0.848 & 1.27\% & 0.327 & 0.028 & 8.56\% & 67.754 & 0.538 & 0.79\% & 0.310 & 0.020 & 6.45\% \\
 & 600  & 67.575 & 0.732 & 1.08\% & 0.313 & 0.023 & 7.35\% & 67.737 & 0.404 & 0.60\% & 0.309 & 0.016 & 5.18\% \\
 & 800  & 67.683 & 0.591 & 0.87\% & 0.319 & 0.019 & 5.96\% & 67.681 & 0.341 & 0.50\% & 0.310 & 0.013 & 4.31\% \\
 &1000  & 67.793 & 0.499 & 0.74\% & 0.316 & 0.016 & 5.06\% & 67.688 & 0.306 & 0.45\% & 0.306 & 0.012 & 3.92\% \\
\hline
\end{tabular}
\end{table}

\begin{table}[!htbp]
\scriptsize
\setlength{\tabcolsep}{2pt}
\renewcommand{\arraystretch}{1.0}
\caption{Summary of the inferred cosmological parameters as a function of the number of detected events $N$, for three different mass ratios $q = 0.455$, $0.636$, and $0.818$. For each $q$, we report the inferred Hubble constant $H_0$, its absolute uncertainty $\sigma_{H_0}$ and relative error ${\cal E}(H_0)$, as well as the matter density parameter $\Omega_m$ with its corresponding uncertainty $\sigma_{\Omega_m}$ and relative error ${\cal E}(\Omega_m)$.}
\label{tab:q}
\begin{tabular}{|c|cccccc|cccccc|cccccc|}
\hline
\multirow{2}{*}{$N$} &
\multicolumn{6}{c|}{q = 0.455} &
\multicolumn{6}{c|}{q = 0.636} &
\multicolumn{6}{c|}{q = 0.818} \\
 & $H_0$ & $\sigma_{H_0}$ & ${\cal E}({H_0})$ & $\Omega_m$ & $\sigma_{\Omega_m}$ & ${\cal E}({\Omega_m})$
 & $H_0$ & $\sigma_{H_0}$ &$ {\cal E}({H_0})$ & $\Omega_m$ & $\sigma_{\Omega_m}$ & ${\cal E}({\Omega_m})$
 & $H_0$ & $\sigma_{H_0}$ & ${\cal E}({H_0})$ & $\Omega_m$ & $\sigma_{\Omega_m}$ & ${\cal E}({\Omega_m})$ \\
\hline
40 & 75.99 & 5.03 & 6.62\% & 0.101 & 0.116 & 115.81\%
   & 69.95 & 5.34 & 7.63\% & 0.248 & 0.166 & 66.85\%
   & 65.30 & 4.72 & 7.22\% & 0.328 & 0.173 & 52.85\% \\
80 & 66.14 & 5.41 & 8.18\% & 0.360 & 0.202 & 56.14\%
   & 67.80 & 4.04 & 5.96\% & 0.299 & 0.141 & 47.13\%
   & 68.80 & 3.57 & 5.19\% & 0.245 & 0.105 & 42.71\% \\
120 & 66.86 & 3.69 & 5.53\% & 0.342 & 0.128 & 37.57\%
    & 69.06 & 3.27 & 4.74\% & 0.261 & 0.098 & 37.64\%
    & 69.14 & 3.32 & 4.80\% & 0.243 & 0.098 & 40.31\% \\
160 & 68.62 & 2.54 & 3.70\% & 0.294 & 0.080 & 27.19\%
    & 67.38 & 2.38 & 3.53\% & 0.333 & 0.084 & 25.21\%
    & 67.29 & 2.34 & 3.48\% & 0.313 & 0.083 & 26.59\% \\
200 & 69.07 & 2.32 & 3.36\% & 0.288 & 0.071 & 24.74\%
    & 68.06 & 2.15 & 3.16\% & 0.310 & 0.073 & 23.38\%
    & 68.11 & 2.15 & 3.15\% & 0.291 & 0.071 & 24.33\% \\
240 & 68.74 & 2.08 & 3.03\% & 0.294 & 0.066 & 22.55\%
    & 67.87 & 2.18 & 3.22\% & 0.314 & 0.074 & 23.72\%
    & 67.46 & 1.94 & 2.87\% & 0.312 & 0.067 & 21.37\% \\
280 & 70.84 & 1.83 & 2.58\% & 0.233 & 0.052 & 22.48\%
    & 68.06 & 2.05 & 3.02\% & 0.307 & 0.070 & 22.67\%
    & 69.09 & 1.90 & 2.75\% & 0.268 & 0.060 & 22.37\% \\
320 & 68.36 & 1.66 & 2.43\% & 0.297 & 0.055 & 18.41\%
    & 67.29 & 1.92 & 2.85\% & 0.333 & 0.066 & 19.95\%
    & 69.27 & 1.76 & 2.54\% & 0.261 & 0.054 & 20.53\% \\
360 & 68.23 & 1.48 & 2.17\% & 0.299 & 0.050 & 16.73\%
    & 67.29 & 1.77 & 2.63\% & 0.327 & 0.060 & 18.39\%
    & 69.16 & 1.53 & 2.22\% & 0.269 & 0.048 & 17.79\% \\
400 & 68.03 & 1.24 & 1.83\% & 0.305 & 0.042 & 13.79\%
    & 68.13 & 1.65 & 2.42\% & 0.294 & 0.053 & 18.19\%
    & 68.71 & 1.43 & 2.08\% & 0.282 & 0.046 & 16.29\% \\
600 & 68.48 & 1.06 & 1.55\% & 0.299 & 0.035 & 11.80\%
    & 67.92 & 1.25 & 1.84\% & 0.297 & 0.041 & 13.76\%
    & 68.40 & 1.11 & 1.62\% & 0.290 & 0.036 & 12.57\% \\
800 & 68.22 & 0.97 & 1.42\% & 0.313 & 0.033 & 10.56\%
    & 67.94 & 1.08 & 1.58\% & 0.305 & 0.036 & 11.72\%
    & 68.33 & 0.99 & 1.45\% & 0.289 & 0.032 & 11.11\% \\
1000 & 68.01 & 0.96 & 1.41\% & 0.320 & 0.032 & 10.09\%
     & 68.00 & 1.00 & 1.48\% & 0.300 & 0.033 & 10.88\%
     & 68.14 & 0.89 & 1.31\% & 0.291 & 0.029 & 9.90\% \\
\hline
\end{tabular}
\end{table}

\vspace{1cm}

\newpage

\section{Conclusions}

In this work, we have investigated what is the impact of the astrophysical parameters of binary neutron star (BNS) on the constraints of the cosmological parameters. From the astrophysical point of view we focused on parameters like the mass ratio distribution ($q$), inclination angle ($\iota$) and the maximum redshift ($z_{\max}$), while the Hubble parameter ($H_0$) and the matter density parameter ($\Omega_m$) where the cosmological counterpart.

Our primary findings demonstrate that the underlying astrophysical assumptions in population synthesis models play a pivotal role in the final precision of standard siren cosmology. While the precision of the Hubble constant ($H_0$) is heavily driven by the absolute sample size and local calibration, constraining the matter density parameter ($\Omega_m$) inherently requires deep-redshift catalogs ($z_{\max} \ge 1.0$). We verified that variations in the simulated mass ratio $q$, often considered a secondary factor in low-redshift approximations, can introduce subtle systematic shifts in the simulated signal-to-noise ratio (SNR) distribution. When combined with the well-known distance--inclination degeneracy, these variations directly affect the Fisher matrix volume and the geometry of the resulting credible regions.

As we transition into the era of third-generation gravitational-wave observatories, represented by the Einstein Telescope (ET) and Cosmic Explorer (CE), the statistical errors on cosmological parameters are expected to shrink to the sub-percent level. Consequently, systematic uncertainties arising from waveform modeling choices, such as IMRPhenomD variations, and prior selection criteria will shift from being negligible to becoming the dominant source of error. Our analysis highlights that forecasting the performance of future networks as precision cosmological probes requires rigorous attention to these underlying astrophysical and phenomenological assumptions.

In conclusion, this study emphasizes that proper modeling of the mass ratio and selection effects is crucial to unlock the full potential of multi-messenger astronomy as an independent cosmological probe. Future extensions of this work will incorporate the impacts of binary precession, tidal deformability parameters, and non-Gaussian lens magnification distributions into the computational analysis to further refine the synergy between gravitational-wave data and fundamental cosmology.

\vspace{1cm}

\begin{acknowledgments}
RSG acknowledges financial support from FAPERJ under Grant No. 260003/005977/2024 (APQ1). GBFL acknowledges financial support from the Coordenação de Aperfeiçoamento de Pessoal de Nível Superior -- Brasil (CAPES), Grant No. 88887.939874/2024-00 (Finance Code 001).
\end{acknowledgments}
\newpage

\section{Bibliography}

\bibliographystyle{unsrt}
\bibliography{references}

@article{Abbott:2016blz,
  author       = "Abbott, B. P. and others",
  collaboration = "LIGO Scientific Collaboration and Virgo Collaboration",
  title        = "{Observation of Gravitational Waves from a Binary Black Hole Merger}",
  journal      = "Phys. Rev. Lett.",
  volume       = "116",
  year         = "2016",
  number       = "6",
  pages        = "061102",
  doi          = "10.1103/PhysRevLett.116.061102",
  eprint       = "1602.03837",
  archivePrefix= "arXiv",
  primaryClass = "gr-qc"
}

@article{Abbott:2019yzh,
  author        = "Abbott, B. P. and others",
  collaboration = "LIGO Scientific Collaboration and Virgo Collaboration",
  title         = "{A Gravitational-wave Measurement of the Hubble Constant Following the Second Observing Run of Advanced LIGO and Virgo}",
  journal       = "Astrophys. J.",
  volume        = "909",
  year          = "2021",
  number        = "2",
  pages         = "218",
  doi           = "10.3847/1538-4357/abdcb7",
  eprint        = "1908.06060",
  archivePrefix = "arXiv",
  primaryClass  = "astro-ph.CO"
}

@article{Abbott:2021djp,
  author        = "Abbott, R. and others",
  collaboration = "LIGO Scientific Collaboration and Virgo Collaboration and KAGRA Collaboration",
  title         = "{GWTC-3: Compact Binary Coalescences Observed by LIGO and Virgo During the Second Part of the Third Observing Run}",
  journal       = "Phys. Rev. X",
  volume        = "13",
  year          = "2023",
  number        = "1",
  pages         = "011048",
  doi           = "10.1103/PhysRevX.13.011048",
  eprint        = "2111.03606",
  archivePrefix = "arXiv",
  primaryClass  = "gr-qc"
}

@article{Muttoni_2023_dark_sirens,
  title = {Dark siren cosmology with binary black holes in the era of third-generation gravitational wave detectors},
  author = {Muttoni, Niccol{\`o} and Laghi, Danny and Tamanini, Nicola and Marsat, Sylvain and Izquierdo-Villalba, David},
  journal = {arXiv:2303.10693},
  year = {2023},
  eprint = {2303.10693},
  archivePrefix = {arXiv},
  primaryClass = {astro-ph.CO}
}

@article{Alfradique_2023_dark_siren_measurement,
  title = {A dark siren measurement of the Hubble constant using gravitational wave events from the first three LIGO/Virgo observing runs and DELVE},
  author = {Alfradique, V. and Bom, C. R. and Palmese, A. and Teixeira, G. and Santana-Silva, L. and others},
  journal = {arXiv:2310.13695},
  year = {2023},
  eprint = {2310.13695},
  archivePrefix = {arXiv},
  primaryClass = {astro-ph.CO}
}

@article{Ezquiaga_Holz_2022_spectral_sirens,
  title = {Spectral sirens: cosmology from the full mass distribution of compact binaries},
  author = {Ezquiaga, Jose Mar{\'\i}a and Holz, Daniel E.},
  journal = {Phys. Rev. Lett.},
  volume = {129},
  number = {6},
  pages = {061102},
  year = {2022},
  doi = {10.1103/PhysRevLett.129.061102},
  eprint = {2202.08240},
  archivePrefix = {arXiv},
  primaryClass = {astro-ph.CO}
}

@article{Farah_2024_spectral_sirens,
  title = {No need to know: astrophysics-free gravitational-wave cosmology},
  author = {Farah, Amanda M. and Callister, Thomas A. and Ezquiaga, Jose Mar{\'\i}a and Zevin, Michael and Holz, Daniel E.},
  journal = {arXiv:2404.02210},
  year = {2024},
  eprint = {2404.02210},
  archivePrefix = {arXiv},
  primaryClass = {astro-ph.CO}
}

@article{Schutz1986_HubbleGW,
  author = {Schutz, B. F.},
  title = {Determining the Hubble constant from gravitational wave observations},
  journal = {Nature},
  volume = {323},
  pages = {310--311},
  year = {1986},
  doi = {10.1038/323310a0}
}

@article{Souza_Sturani_2021_ET_cosmography,
  title = {Cosmography with standard sirens from the Einstein Telescope},
  author = {de Souza, Josiel Mendonça Soares and Sturani, Riccardo and Alcaniz, Jailson},
  journal = {arXiv:2110.13316},
  year = {2021},
  eprint = {2110.13316},
  archivePrefix = {arXiv},
  primaryClass = {astro-ph.CO}
}

@article{Zhang2019_ET_standard_sirens,
  title = {Cosmological parameter estimation with future gravitational wave standard siren observation from the Einstein Telescope},
  author = {Zhang, Jing-Fei and Zhang, Ming and Jin, Shang-Jie and Qi, Jing-Zhao and Zhang, Xin},
  journal = {arXiv:1907.03238},
  year = {2019},
  eprint = {1907.03238},
  archivePrefix = {arXiv},
  primaryClass = {astro-ph.CO}
}

@article{Abbott2017_GW170817,
  title = {GW170817: Observation of Gravitational Waves from a Binary Neutron Star Inspiral},
  author = {B. P. Abbott et al. (LIGO Scientific Collaboration and Virgo Collaboration)},
  journal = {Phys. Rev. Lett.},
  volume = {119},
  pages = {161101},
  year = {2017},
  doi = {10.1103/PhysRevLett.119.161101},
  eprint = {1710.05832},
  archivePrefix = {arXiv},
  primaryClass = {gr-qc}
}

@article{Abbott2017_H0_GW170817,
  title = {A gravitational-wave standard siren measurement of the Hubble constant},
  author = {B. P. Abbott et al. (LIGO Scientific Collaboration and Virgo Collaboration and others)},
  journal = {Nature},
  volume = {551},
  pages = {85--88},
  year = {2017},
  doi = {10.1038/nature24471},
  eprint = {1710.05835},
  archivePrefix = {arXiv},
  primaryClass = {astro-ph.CO}
}

@article{Pretorius2005,
  title={Evolution of binary black-hole spacetimes},
  author={Pretorius, Frans},
  journal={Phys. Rev. Lett.},
  volume={95},
  pages={121101},
  year={2005},
  doi={10.1103/PhysRevLett.95.121101}
}

@article{Husa2016_IMRPhenomI,
  title = {Frequency-domain gravitational waves from non-precessing black-hole binaries. I. New numerical waveforms and anatomy of the signal},
  author = {Husa, Sascha and Khan, Sebastian and Hannam, Mark and P\"urrer, Michael and Ohme, Frank and Jim\'enez Forteza, Xisco and Boh\'e, Alejandro},
  journal = {Phys. Rev. D},
  volume = {93},
  pages = {044006},
  year = {2016},
  doi = {10.1103/PhysRevD.93.044006},
  eprint = {1508.07250},
  archivePrefix = {arXiv},
  primaryClass = {gr-qc}
}

@article{Khan2016_IMRPhenomD,
  title = {Frequency-domain gravitational waves from non-precessing black-hole binaries. II. A phenomenological model for the advanced detector era},
  author = {Khan, S. and Husa, S. and Hannam, M. and Ohme, F. and P{\"u}rrer, M. and Jim{\'e}nez Forteza, X. and Boh{\'e}, A.},
  journal = {Phys. Rev. D},
  volume = {93},
  pages = {044007},
  year = {2016},
  doi = {10.1103/PhysRevD.93.044007},
  eprint = {1508.07253},
  archivePrefix = {arXiv},
  primaryClass = {gr-qc}
}

@article{Saleem2019_inclination_distribution,
  title = {Prospects of joint detections of neutron star mergers and short-GRBs with Gaussian structured jets},
  author = {Saleem, M.},
  journal = {Monthly Notices of the Royal Astronomical Society},
  volume = {493},
  pages = {1633},
  year = {2020},
}

@article{Kathirgamaraju_2017,
   title={Off-axis short GRBs from structured jets as counterparts to GW events},
   volume={473},
   ISSN={1745-3933},
   url={http://dx.doi.org/10.1093/mnrasl/slx175},
   DOI={10.1093/mnrasl/slx175},
   number={1},
   journal={Monthly Notices of the Royal Astronomical Society: Letters},
   publisher={Oxford University Press (OUP)},
   author={Kathirgamaraju, Adithan and Barniol Duran, Rodolfo and Giannios, Dimitrios},
   year={2017},
   month=Oct, pages={L121–L125} }

@article{de_Souza_2023,
   title={GWDALI: A Fisher-matrix based software for gravitational wave parameter-estimation beyond Gaussian approximation},
   volume={45},
   ISSN={2213-1337},
   url={http://dx.doi.org/10.1016/j.ascom.2023.100759},
   DOI={10.1016/j.ascom.2023.100759},
   journal={Astronomy and Computing},
   publisher={Elsevier BV},
   author={de Souza, J.M.S. and Sturani, R.},
   year={2023},
   month=oct, pages={100759} }

@article{Dupletsa_2023,
   title={gwfish: A simulation software to evaluate parameter-estimation capabilities of gravitational-wave detector networks},
   volume={42},
   ISSN={2213-1337},
   url={http://dx.doi.org/10.1016/j.ascom.2022.100671},
   DOI={10.1016/j.ascom.2022.100671},
   journal={Astronomy and Computing},
   publisher={Elsevier BV},
   author={Dupletsa, U. and Harms, J. and Banerjee, B. and Branchesi, M. and Goncharov, B. and Maselli, A. and Oliveira, A.C.S. and Ronchini, S. and Tissino, J.},
   year={2023},
   month=jan, pages={100671} }

@article{Zhao_2011,
   title={Determination of dark energy by the Einstein Telescope: Comparing with CMB, BAO, and SNIa observations},
   volume={83},
   ISSN={1550-2368},
   url={http://dx.doi.org/10.1103/PhysRevD.83.023005},
   DOI={10.1103/physrevd.83.023005},
   number={2},
   journal={Physical Review D},
   publisher={American Physical Society (APS)},
   author={Zhao, W. and Van Den Broeck, C. and Baskaran, D. and Li, T. G. F.},
   year={2011},
   month=jan }

@article{2020aganin,
   title={Planck2018 results: VI. Cosmological parameters},
   volume={641},
   ISSN={1432-0746},
   url={http://dx.doi.org/10.1051/0004-6361/201833910},
   DOI={10.1051/0004-6361/201833910},
   journal={Astronomy \&; Astrophysics},
   publisher={EDP Sciences},
   author={Aghanim, N. et al},
   year={2020},
   month=sep, pages={A6} }

@article{Cai_2017,
   title={Estimating cosmological parameters by the simulated data of gravitational waves from the Einstein Telescope},
   volume={95},
   ISSN={2470-0029},
   url={http://dx.doi.org/10.1103/PhysRevD.95.044024},
   DOI={10.1103/physrevd.95.044024},
   number={4},
   journal={Physical Review D},
   publisher={American Physical Society (APS)},
   author={Cai, Rong-Gen and Yang, Tao},
   year={2017},
   month=feb }

@article{Jin_2022,
   title={Impacts of gravitational-wave standard siren observations from Einstein Telescope and Cosmic Explorer on weighing neutrinos in interacting dark energy models},
   volume={74},
   ISSN={1572-9494},
   url={http://dx.doi.org/10.1088/1572-9494/ac7b76},
   DOI={10.1088/1572-9494/ac7b76},
   number={10},
   journal={Communications in Theoretical Physics},
   publisher={IOP Publishing},
   author={Jin, Shang-Jie and Zhu, Rui-Qi and Wang, Ling-Feng and Li, Hai-Li and Zhang, Jing-Fei and Zhang, Xin},
   year={2022},
   month=sep, pages={105404} }

@article{de_Souza_2023_luminosity_distance,
   title={Luminosity distance uncertainties from gravitational wave detections of binary neutron stars by third generation observatories},
   volume={108},
   ISSN={2470-0029},
   url={http://dx.doi.org/10.1103/PhysRevD.108.043027},
   DOI={10.1103/physrevd.108.043027},
   number={4},
   journal={Physical Review D},
   publisher={American Physical Society (APS)},
   author={de Souza, Josiel Mendonça Soares and Sturani, Riccardo},
   year={2023},
   month=aug }

@article{Khan:2015jqa,
    author = {Khan, Sebastian and Husa, Sascha and Hannam, Mark and Ohme, Frank and P{\"u}rrer, Michael and Jim{\'e}nez Forteza, Xisco and Boh{\'e}, Alejandro},
    title = "{Frequency-domain gravitational waves from nonprecessing black-hole binaries. II. A phenomenological model for the advanced detector era}",
    eprint = "1508.07253",
    archivePrefix = "arXiv",
    primaryClass = "gr-qc",
    doi = "10.1103/PhysRevD.93.044007",
    journal = "Phys. Rev. D",
    volume = "93",
    number = "4",
    pages = "044007",
    year = "2016"
}

@article{Ajith:2009bn,
    author = "Ajith, P. and others",
    title = "{Inspiral-merger-ringdown waveforms for black-hole binaries with non-precessing spins}",
    eprint = "0909.2867",
    archivePrefix = "arXiv",
    primaryClass = "gr-qc",
    doi = "10.1103/PhysRevLett.106.241101",
    journal = "Phys. Rev. Lett.",
    volume = "106",
    pages = "241101",
    year = "2011"
}

@article{Taracchini:2012ig,
    author = "Taracchini, Andrea and Pan, Yi and Buonanno, Alessandra and Barausse, Enrico and Boyle, Michael and Chu, Tony and Lovelace, Geoffrey and Pfeiffer, Harald P. and Scheel, Mark A.",
    title = "{Prototype effective-one-body model for nonprecessing spinning inspiral-merger-ringdown waveforms}",
    eprint = "1202.0790",
    archivePrefix = "arXiv",
    primaryClass = "gr-qc",
    doi = "10.1103/PhysRevD.86.024011",
    journal = "Phys. Rev. D",
    volume = "86",
    pages = "024011",
    year = "2012"
}

@article{Buonanno:2009zt,
    author = "Buonanno, Alessandra and Iyer, Bala and Ochsner, Evan and Pan, Yi and Sathyaprakash, B. S.",
    title = "{Comparison of post-Newtonian templates for compact binary inspiral signals in gravitational-wave detectors}",
    eprint = "0907.0700",
    archivePrefix = "arXiv",
    primaryClass = "gr-qc",
    doi = "10.1103/PhysRevD.80.084043",
    journal = "Phys. Rev. D",
    volume = "80",
    pages = "084043",
    year = "2009"
}

@article{Santamaria:2010yb,
    author = "Santamaria, L. and others",
    title = "{Matching post-Newtonian and numerical relativity waveforms: systematic errors and a new phenomenological model for non-precessing black hole binaries}",
    eprint = "1005.3306",
    archivePrefix = "arXiv",
    primaryClass = "gr-qc",
    reportNumber = "LIGO-P1000048, AEI-2010-122",
    doi = "10.1103/PhysRevD.82.064016",
    journal = "Phys. Rev. D",
    volume = "82",
    pages = "064016",
    year = "2010"
}

@article{Ghirlanda_2019,
   title={Compact radio emission indicates a structured jet was produced by a binary neutron star merger},
   volume={363},
   ISSN={1095-9203},
   url={http://dx.doi.org/10.1126/science.aau8815},
   DOI={10.1126/science.aau8815},
   number={6430},
   journal={Science},
   publisher={American Association for the Advancement of Science (AAAS)},
   author={Ghirlanda, G. and Salafia, O. S. and Paragi, Z. and Giroletti, M. and Yang, J. and Marcote, B. and Blanchard, J. and Agudo, I. and An, T. and Bernardini, M. G. and Beswick, R. and Branchesi, M. and Campana, S. and Casadio, C. and Chassande-Mottin, E. and Colpi, M. and Covino, S. and D’Avanzo, P. and D’Elia, V. and Frey, S. and Gawronski, M. and Ghisellini, G. and Gurvits, L. I. and Jonker, P. G. and van Langevelde, H. J. and Melandri, A. and Moldon, J. and Nava, L. and Perego, A. and Perez-Torres, M. A. and Reynolds, C. and Salvaterra, R. and Tagliaferri, G. and Venturi, T. and Vergani, S. D. and Zhang, M.},
   year={2019},
   month=Mar, pages={968–971} }

@article{Mooley_2018,
   title={Superluminal motion of a relativistic jet in the neutron-star merger GW170817},
   volume={561},
   ISSN={1476-4687},
   url={http://dx.doi.org/10.1038/s41586-018-0486-3},
   DOI={10.1038/s41586-018-0486-3},
   number={7723},
   journal={Nature},
   publisher={Springer Science and Business Media LLC},
   author={Mooley, K. P. and Deller, A. T. and Gottlieb, O. and Nakar, E. and Hallinan, G. and Bourke, S. and Frail, D. A. and Horesh, A. and Corsi, A. and Hotokezaka, K.},
   year={2018},
   month=sep, pages={355–359} }

@article{taylorTF2-1,
    author  = {Damour, T. and Iyer, B. R. and Sathyaprakash, B. S.},
    title   = {Comparison of search templates for gravitational waves from binary inspiral},
    journal = {Phys. Rev. D},
    volume  = {63},
    pages   = {044023},
    year    = {2001},
    doi     = {10.1103/PhysRevD.63.044023}
}

@article{taylorTF2-2,
    author  = {Damour, T. and Iyer, B. R. and Sathyaprakash, B. S.},
    title   = {Comparison of search templates for gravitational waves from binary inspiral: 3.5PN update},
    journal = {Phys. Rev. D},
    volume  = {66},
    pages   = {027502},
    year    = {2002},
    doi     = {10.1103/PhysRevD.66.027502}
}

@article{taylorTF2-3,
    author  = {Damour, T. and Iyer, B. R. and Sathyaprakash, B. S.},
    title   = {Erratum: Comparison of search templates for gravitational waves from binary inspiral: 3.5PN update},
    journal = {Phys. Rev. D},
    volume  = {72},
    pages   = {029901},
    year    = {2005},
    doi     = {10.1103/PhysRevD.72.029901}
}

@article{taylorTF2-4,
    author  = {Arun, K. G. and Iyer, B. R. and Sathyaprakash, B. S. and Sundararajan, P. A.},
    title   = {Parameter estimation of inspiralling compact binaries using 3.5 post-Newtonian gravitational wave phasing: The nonspinning case},
    journal = {Phys. Rev. D},
    volume  = {71},
    pages   = {084008},
    year    = {2005},
    doi     = {10.1103/PhysRevD.71.084008}
}

@misc{sxs,
    title        = {{SXS Catalog of Simulations}},
    howpublished = {\url{https://data.black-holes.org/simulations/index.html}},
    note         = {SXS Collaboration}
}

@article{bam1,
    author  = {Bruegmann, Bernd and Gonzalez, Jose A. and Hannam, Mark and Husa, Sascha and Sperhake, Ulrich and Tichy, Wolfgang},
    title   = {Calibration of moving puncture simulations},
    journal = {Phys. Rev. D},
    volume  = {77},
    pages   = {024027},
    year    = {2008},
    doi     = {10.1103/PhysRevD.77.024027}
}

@article{bam2,
    author  = {Husa, Sascha and Gonzalez, Jose A. and Hannam, Mark and Bruegmann, Bernd and Sperhake, Ulrich},
    title   = {Reducing phase error in long numerical binary black hole evolutions with sixth-order finite differencing},
    journal = {Class. Quantum Grav.},
    volume  = {25},
    pages   = {105006},
    year    = {2008},
    doi     = {10.1088/0264-9381/25/10/105006}
}

@article{taracchini,
    author  = {Taracchini, Andrea and Buonanno, Alessandra and Pan, Yi and Hinderer, Tanja and Boyle, Michael and Hemberger, Daniel A. and Kidder, Lawrence E. and Lovelace, Geoffrey and Mroue, Abdul H. and Pfeiffer, Harald P. and Scheel, Mark A. and Szilagyi, Bela and Taylor, Nicholas W. and Zenginoglu, Anil},
    title   = {Effective-one-body model for black-hole binaries with generic mass ratios and spins},
    journal = {Phys. Rev. D},
    volume  = {89},
    pages   = {061502},
    year    = {2014},
    doi     = {10.1103/PhysRevD.89.061502}
}

@article{hyb1,
    author  = {Ajith, P. and Babak, S. and Chen, Y. and Hewitson, M. and Krishnan, B. and Sintes, A. M. and Whelan, J. T. and Bruegmann, B. and Diener, P. and Dorband, N. and Gonzalez, J. and Hannam, M. and Husa, S. and Pollney, D. and Rezzolla, L. and Santamaria, L. and Sperhake, U. and Thornburg, J.},
    title   = {Template bank for gravitational waveforms from coalescing binary black holes: Nonspinning binaries},
    journal = {Phys. Rev. D},
    volume  = {77},
    pages   = {104017},
    year    = {2008},
    doi     = {10.1103/PhysRevD.77.104017}
}

@article{hyb2,
    author  = {MacDonald, Ilana and Nissanke, Samaya and Pfeiffer, Harald P.},
    title   = {Suitability of post-Newtonian/numerical-relativity hybrid waveforms for gravitational wave detectors},
    journal = {Class. Quantum Grav.},
    volume  = {28},
    pages   = {134002},
    year    = {2011},
    doi     = {10.1088/0264-9381/28/13/134002}
}

@article{hyb3,
    author  = {Boyle, Michael},
    title   = {Uncertainty in hybrid gravitational waveforms: Optimizing initial orbital frequencies for binary black-hole simulations},
    journal = {Phys. Rev. D},
    volume  = {84},
    pages   = {064013},
    year    = {2011},
    doi     = {10.1103/PhysRevD.84.064013}
}

@article{hyb4,
    author  = {Ajith, P. and Boyle, M. and Brown, D. A. and Bruegmann, B. and Buchman, L. T. and Cadonati, L. and others},
    title   = {The {NINJA-2} catalog of hybrid post-Newtonian/numerical-relativity waveforms for non-precessing black-hole binaries},
    journal = {Class. Quantum Grav.},
    volume  = {29},
    pages   = {124001},
    year    = {2012},
    doi     = {10.1088/0264-9381/29/12/124001}
}

@article{Dietrich_2017,
   title={Closed-form tidal approximants for binary neutron star gravitational waveforms constructed from high-resolution numerical relativity simulations},
   volume={96},
   ISSN={2470-0029},
   url={http://dx.doi.org/10.1103/PhysRevD.96.121501},
   DOI={10.1103/physrevd.96.121501},
   number={12},
   journal={Physical Review D},
   publisher={American Physical Society (APS)},
   author={Dietrich, Tim and Bernuzzi, Sebastiano and Tichy, Wolfgang},
   year={2017},
   month=Dec }

@article{De_2018,
   title={Tidal Deformabilities and Radii of Neutron Stars from the Observation of GW170817},
   volume={121},
   ISSN={1079-7114},
   url={http://dx.doi.org/10.1103/PhysRevLett.121.091102},
   DOI={10.1103/physrevlett.121.091102},
   number={9},
   journal={Physical Review Letters},
   publisher={American Physical Society (APS)},
   author={De, Soumi and Finstad, Daniel and Lattimer, James M. and Brown, Duncan A. and Berger, Edo and Biwer, Christopher M.},
   year={2018},
   month=Aug }

@article{Hinderer_2008,
    author = {Hinderer, Tanja},
    title = {Tidal Love Numbers of Neutron Stars},
    journal = {The Astrophysical Journal},
    volume = {677},
    number = {2},
    pages = {1216--1220},
    year = {2008},
    doi = {10.1086/533487},
    eprint = {0711.2420},
    archivePrefix = {arXiv},
    primaryClass = {astro-ph}
}

@misc{LVK_O4_250,
  author       = {{LIGO Scientific Collaboration}},
  title        = {{LIGO--Virgo--KAGRA Complete Fourth Observing Run}},
  year         = {2025},
  month        = nov,
  howpublished = {\url{https://www.ligo.caltech.edu/WA/news/ligo20251118}}
}

\end{document}